\documentclass[]{spie}  

\usepackage{amsmath,amsfonts,amssymb}
\usepackage{graphicx}
\usepackage[colorlinks=true, allcolors=blue]{hyperref}

\title{The PLACID active coronagraphic imager instrument: Commissioning status}

\author[*a]{Ruben Tandon}
\author[a]{Lucas Marquis}
\author[a]{Liurong Lin}
\author[b]{Derya \"Ozt\"urk \c{C}etni}
\author[c]{Axel Potier}
\author[d]{Audrey Baur}
\author[a]{Iljadin Manurung}
\author[a]{Matthias Br\"andli}
\author[a]{Martin Rieder}
\author[a]{Daniele Piazza}
\author[d]{Laurent Jolissaint}
\author[a]{Jonas G. K\"uhn}

\affil[a]{Department of Space and Planetary Sciences, University of Bern, Gesellschaftsstrasse 6, 3012 Bern, Switzerland}
\affil[b]{T\"urkiye National Observatories, Atat\"urk University District, Prof. Dr. L\"utf\"u \"Ulk\"umen Street 8/3, 25050 Yakutiye/Erzurum, Turkey}
\affil[c]{LIRA, Université Paris Cité, Observatoire de Paris, Université PSL, Sorbonne Université, CNRS, F-92190 Meudon, France}
\affil[d]{University of Applied Sciences HEIG-VD, Route de Cheseaux 1, 1401 Yverdon-les-Bains, Switzerland}

\authorinfo{Further author information: \\ *Ruben Tandon: E-mail: ruben.tandon@unibe.ch, Telephone: +41 31 684 32 90}

\begin{document} 
\maketitle

\begin{abstract}
The world’s first adaptive stellar coronagraph, the Programmable Liquid-crystal Active Coronagraphic Imager for the 4-m DAG telescope (PLACID) uses a spatial light modulator operating from H- to Ks-band to dynamically adjust the focal-plane phase mask entirely in software. Positioned between the TROIA XAO system and the DIRAC infrared detector, PLACID was installed on the Nasmyth platform of the Turkish 4-m DAG telescope in 2025, followed by optical alignment and successful preliminary acceptance tests with the calibration light source in early 2026. On-sky commissioning is expected, as soon as the TROIA XAO system will be operational. When on-sky, it will enable high-contrast imaging of exoplanets, brown dwarfs, disks, and binary systems, being able to easily deploy any desired phase pattern in the focal plane. Upcoming features include self-calibration of non-common path aberrations, coronagraphic nulling of binary-stars, and coherent differential imaging. We present the preliminary acceptance procedures and internal alignment results, discovery-space estimates, new binary star features and observation tools, to be ready for first science by late 2026/early 2027.
\end{abstract}


\section{INTRODUCTION}
\label{sec:intro}  

The Programmable Liquid-crystal Active Coronagraphic Imager for the DAG telescope (PLACID) uses a customized spatial light modulator (SLM) in the focal plane, to dynamically adjust the coronagraphic focal-plane phase mask (FPM) in real time to changing observing conditions or science case, thereby distinguishing it as the world’s first “adaptive” exoplanet imaging coronagraph instrument deployed on sky. \\
With just the click of a button, PLACID can switch from one coronagraphic FPM pattern (or any custom crafted phase mask, for instance) to the other. The pixelated FPM can be modified or shifted entirely via software, free of any mechanical actuation. This enables observers to select from a wide range of coronagraphic modes and tailor observations to specific science requirements, or instrument-specific optical conditions, such as non-common path aberrations (NCPAs). PLACID’s main science objective is to perform direct imaging of exoplanets, brown dwarfs, binaries, and circumstellar disks in the Northern hemisphere, operating in the near-infrared, specifically the H- to Ks-bands. \\

The idea of “adaptive coronagraphy” originates from a paper by Bourget et al. in 2001, using a Lyot coronagraph with tunable diameter occulting spot made of mercury \cite{2001PASP..113..436B}. The suggestion of using liquid-crystal on-silicon spatial light modulators (LCOS SLMs, see e.g. Zhang et al. 2014 \cite{2014LSA.....3.e213Z}) as active programmable FPM coronagraphs was introduced by Kühn et al. 2016 \cite{10.1117/12.2232660}. This approach takes advantage of the SLM resolution often exceeding 1 Mpixel and $\sim 10 \mu$m pixel pitch, allowing for sufficient sampling of the telescope point-spread function (PSF) in the focal plane \cite{2025SPIE13627E..0ZK}. However, SLMs imprint a scalar phase retardance, hence limiting the broadband contrast performance - and classical SLMs also require linearly-polarized input light to achieve phase modulation, thus requiring the use of a linear polarizer, limiting optical transmission to less than 50\% for unpolarized light \cite{2025SPIE13627E..0ZK}. Nevertheless, post-adaptive optics (AO) wavefront error (WFE) residuals and leakage from the central obstruction caused by the telescope secondary mirror are the dominant contributors to the leakage budget, which would mean that the aforementioned limitations may not represent an obstacle to contrast from the ground \cite{2022ApOpt..61.9000K}. \\
On the other hand, an SLM-based programmable FPM coronagraph would allow astronomers to adapt to observing conditions, such as seeing and post-AO residual WFE profile, in real time by having the freedom of choosing an optimal FPM within a trade off between inner working angle (IWA) and robustness to low-order aberrations (in particular tip/tilt jitter). The configuration can also be adapted to the specifics of the science target, for example whether the target star is resolved or unresolved, (relevant to ELTs in the future) or whether the target is part of a multiple-star system \cite{2025SPIE13627E..0ZK}. \\
The nulling of several stars within the field of view (FOV) of the instrument is largely facilitated by the implementation of a programmable FPM, enabling niche high-contrast science cases, such as challenging compact binary or triple-star systems of similar magnitudes, instead of having to mechanically position two manufactured phase masks in cascade \cite{10.1117/12.2232660, 2018SPIE10702E..42K}. The SLM focal plane phase pattern can be updated at rates of 30 Hz or higher \cite{2025SPIE13627E..0ZK}, which would enable a multiple-star high contrast imaging mode compatible with field rotation of the altitude-azimuth telescope (with activated pupil-tracking), i.e. post-processing with Angular Differential Imaging (ADI) \cite{2006ApJ...641..556M}. This feature would open a key niche science case for the instrument, by enabling to look for circumbinary companions or disks (see section \ref{sec: binary_mode}). \\
Furthermore, adaptive FPM coronagraphy features include self-calibration of NCPAs, with the SLM generating a Zernike wavefront sensor (WFS) pattern \cite{2016A&A...592A..79N}, as well as phase-shifted variants of it \cite{2012SPIE.8447E..2KW, 2018SPIE10706E..2NK}, without the need of physically moving a component in and out of the optical path. \\ 
Finally, SLMs could be used to introduce local phase shifts at specific modulation frequencies in the focal plane, potentially enabling time-domain coherent differential imaging (CDI) to dynamically discriminate coherent speckles from incoherent astrophysical sources, such as planets \cite{2018SPIE10706E..2NK}.\\ 

\begin{table}[ht]
\caption{Data table for the DAG telescope \cite{2022SPIE12185E..1WK, 2018SPIE10700E..2JY, Author_InPrep}} 
\label{tab:DAG}
\begin{center}       
\begin{tabular}{|l|l|} 
\hline
\rule[-1ex]{0pt}{3.5ex}  \textbf{Name} & \textbf{Do\u{g}u Anadolu Gözlemevi (DAG)}  \\
\hline
\rule[-1ex]{0pt}{3.5ex}  \textbf{Location} & Karakaya Ridge, Erzurum, Turkey   \\
\hline
\rule[-1ex]{0pt}{3.5ex}  \textbf{Latitude} & 39$^{\circ}$46'50.0" N   \\
\hline
\rule[-1ex]{0pt}{3.5ex}  \textbf{Longitude} & 41$^{\circ}$13'36.0" E   \\
\hline
\rule[-1ex]{0pt}{3.5ex}  \textbf{Altitude} & 3170 m   \\
\hline
\rule[-1ex]{0pt}{3.5ex}  \textbf{Seeing (minimum, median, mean)} & 0.24", 1.16", 2.00" \\
\hline
\rule[-1ex]{0pt}{3.5ex}  \textbf{Primary mirror} & 4 m  \\
\hline
\rule[-1ex]{0pt}{3.5ex}  \textbf{Focal length} & 56 m  \\
\hline
\rule[-1ex]{0pt}{3.5ex}  \textbf{Mounting} & Altitude - Azimuth  \\
\hline
\rule[-1ex]{0pt}{3.5ex}  \textbf{Telescope build} & Ritchey - Chrétien  \\
\hline
\rule[-1ex]{0pt}{3.5ex}  \textbf{Declination limit} & $\geq$ -24$^{\circ}$  \\
\hline
\end{tabular}
\end{center}
\end{table}

\begin{figure} [ht]
	\begin{center}
   		\begin{tabular}{c} 
   		\includegraphics[width=\textwidth]{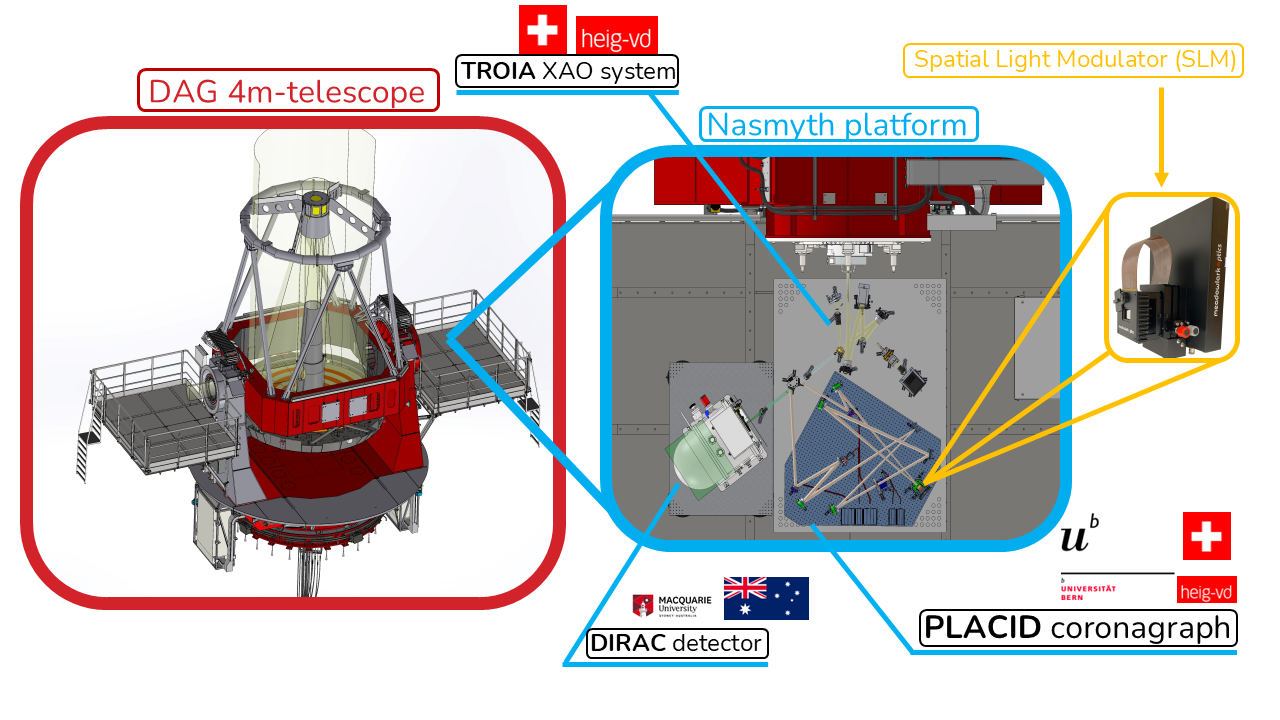}
		\end{tabular}
	\end{center}
   \caption[example] 
   { \label{fig:Nasmyth}The DAG telescope and the instrument setup on the Nasmyth platform (image credit: ATASAM/TNO/DAG observatory, AVR Optics and Meadowlark Optics)}
\end{figure}

The PLACID coronagraph is one of the first instruments of the new Turkish 4-m DAG (Do\u{g}u Anadolu G\"ozlemevi, East Anatolia Observatory, see details in Table \ref{tab:DAG}) observatory and was delivered to the Turkish National Observatories (TNO) facilities at Atatürk University in Erzurum in March 2024. PLACID is placed between two other instruments on a Nasmyth platform - the TuRkish adaptive Optics for Infrared Astronomy (TROIA) \cite{2022SPIE12185E..1WK} extreme adaptive optics (XAO) system and the DAG InfraRed Adaptive optics Camera (DIRAC) with a HAWAII-1RG focal-plane array (see Figures \ref{fig:Nasmyth} and \ref{fig:PLACIDplan}) \cite{2022SPIE12184E..40Z}. The adaptive optics system was developed by HEIG-VD (University of Applied Sciences Western Switzerland, Yverdon-les-Bains, Switzerland), using a 468-actuator ALPAO deformable mirror (DM), a pyramid wavefront sensor (P-WFS), and a low-noise EMCCD WFS-camera. The estimated TROIA residual wavefront errors are expected to be in the nm-range \cite{2022SPIE12185E..1WK}. The PLACID coronagraph can be moved in and out of the NIR science path using a fore-optics intermediate stage platform, located between TROIA and DIRAC. The infrared science camera DIRAC was developed by Australian Astronomical Optics (AAO) at Macquarie University in Sydney, Australia.\\ 
The DAG telescope Assembly, Integration and Validation (AIV) activities on the diffraction-limited Nasmyth have progressed, with initial alignment of the TROIA XAO and the DIRAC infrared detector beam path completed by late 2025. After PLACID Nasmyth commissioning using the internal calibration light source was completed in February 2026, the on-sky commissioning is thus expected to be performed during the second half of 2026. After the successful completion of currently ongoing AO loop closing activities, PLACID first light is anticipated for late summer/early fall of 2026, with science observations to start in late 2026/early 2027. \\

We present in this paper the PLACID alignment and preliminary acceptance measurements undertaken in early 2026, as well as preparations for first science observations, including an update of the discovery space (see section \ref{sec: discovery_space}) and the binary mask mode update of the PLACID graphical user interface (GUI, see section \ref{sec: binary_mode}). The PLACID data reduction pipeline has been developed using the PynPoint python package \cite{2012MNRAS.427..948A}, but will not be detailed in this work.


\section{PLACID PRELIMINARY ACCEPTANCE - ALIGNMENT AND INTERNAL SOURCE TESTS} \label{sec: PA}

\begin{figure} [ht]
	\centering
   \includegraphics[width=0.7\textwidth]{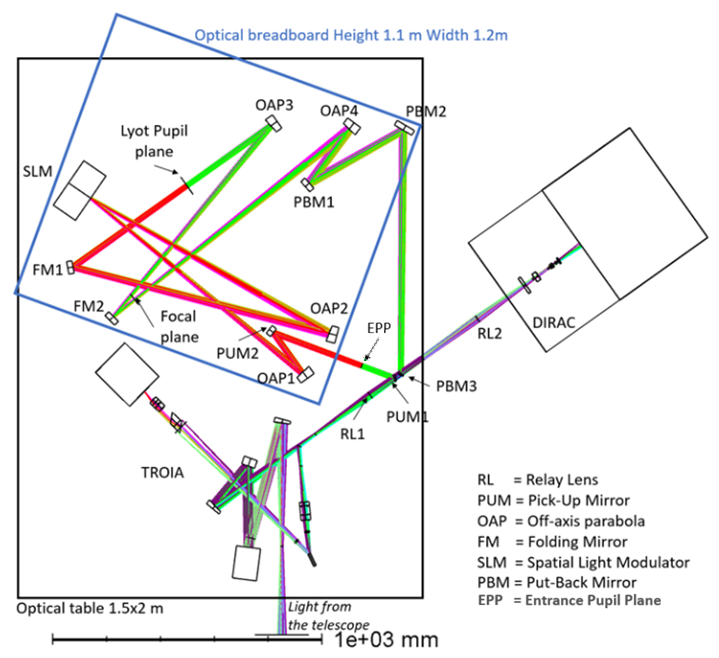}
   \caption[example] 
   { \label{fig:PLACIDplan}Optics plan of PLACID in detail (blue rectangle), and the other two instruments on the Nasmyth, DIRAC and TROIA. \cite{2025SPIE13627E..0ZK}}
\end{figure}

\begin{figure}[htbp]
    \centering
    \begin{tabular}{cc}
        \includegraphics[width=0.45\textwidth]{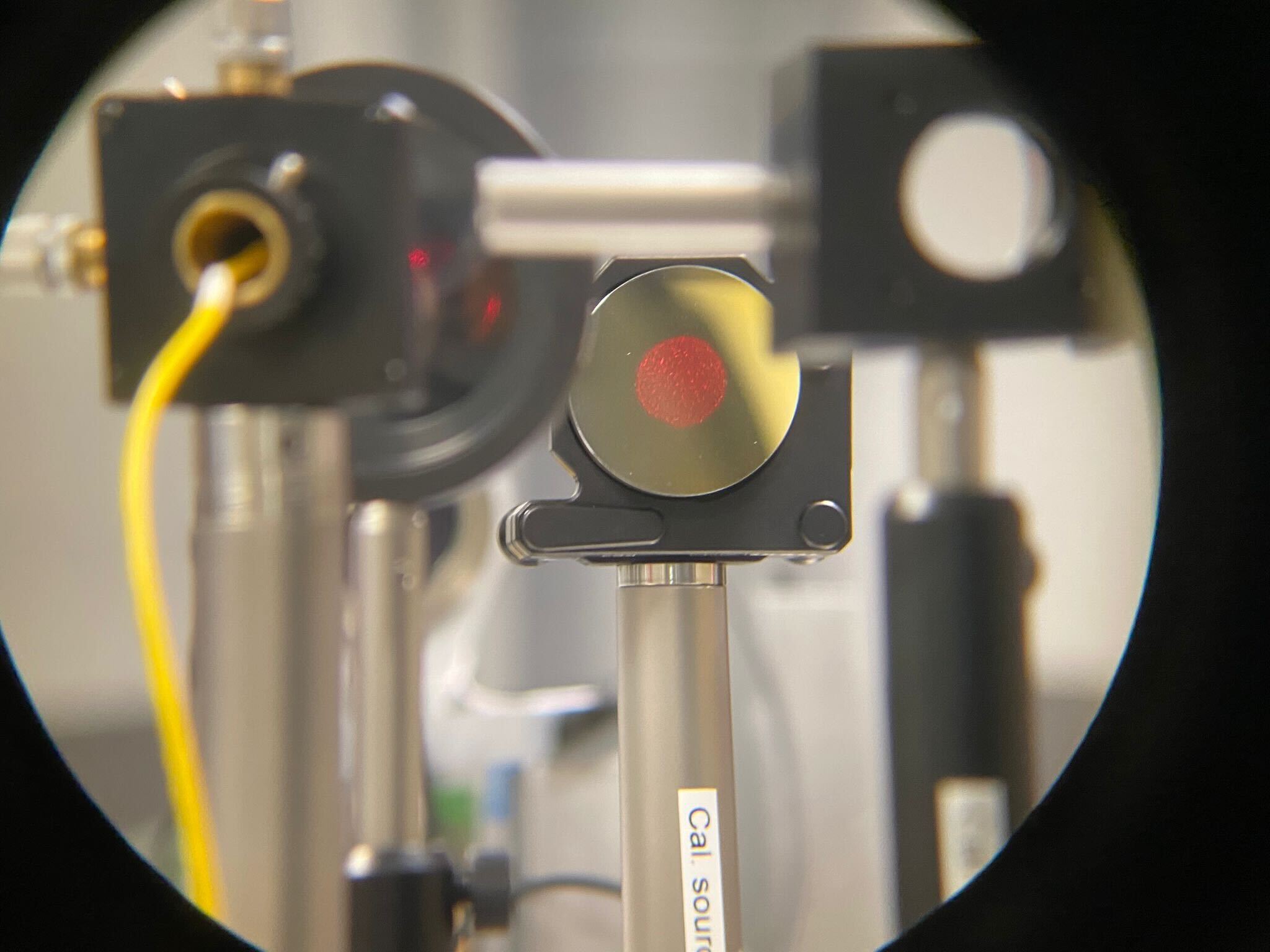} &
        \includegraphics[width=0.45\textwidth]{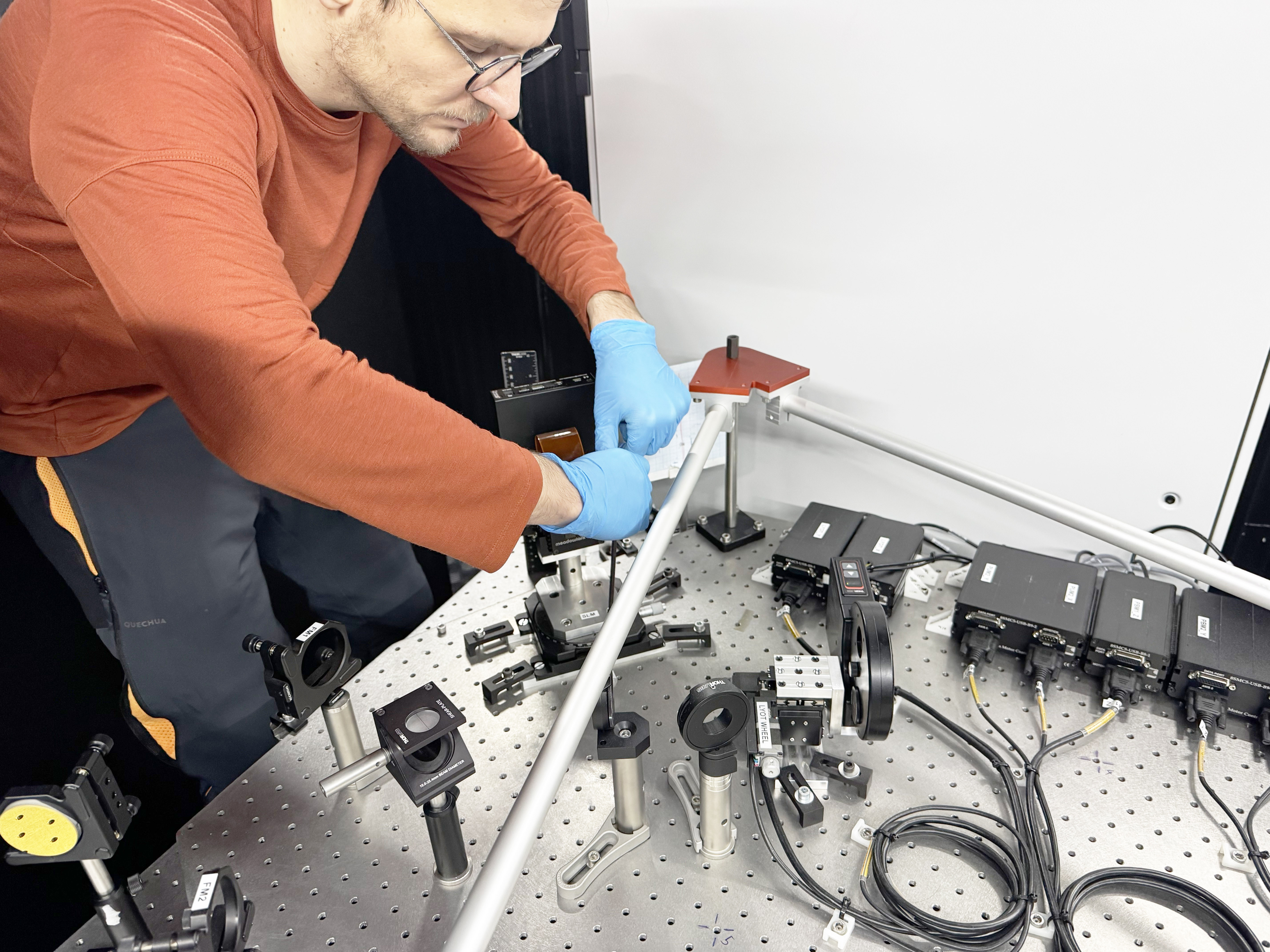} \\

        \includegraphics[width=0.45\textwidth]{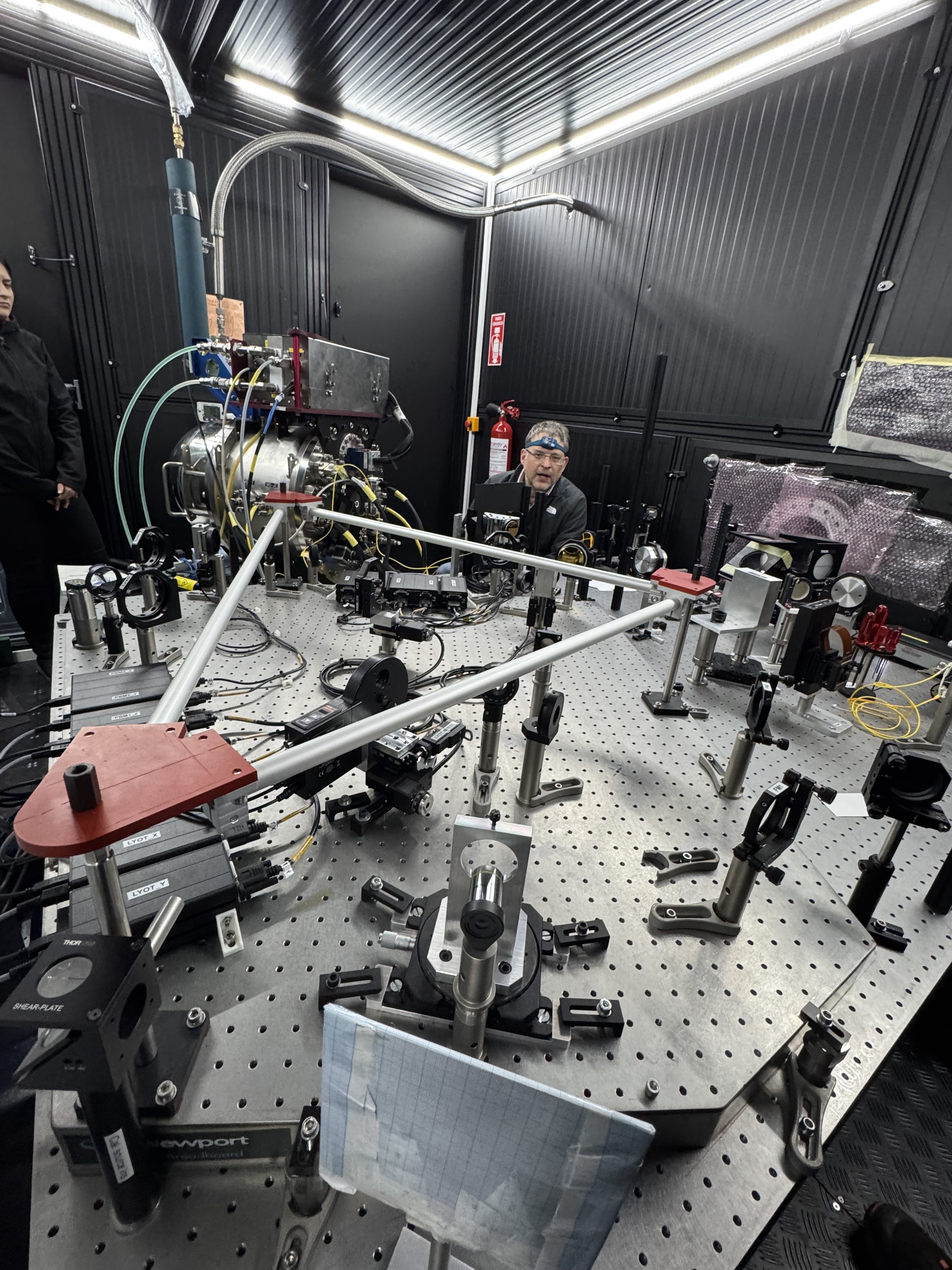} &
        \includegraphics[width=0.45\textwidth]{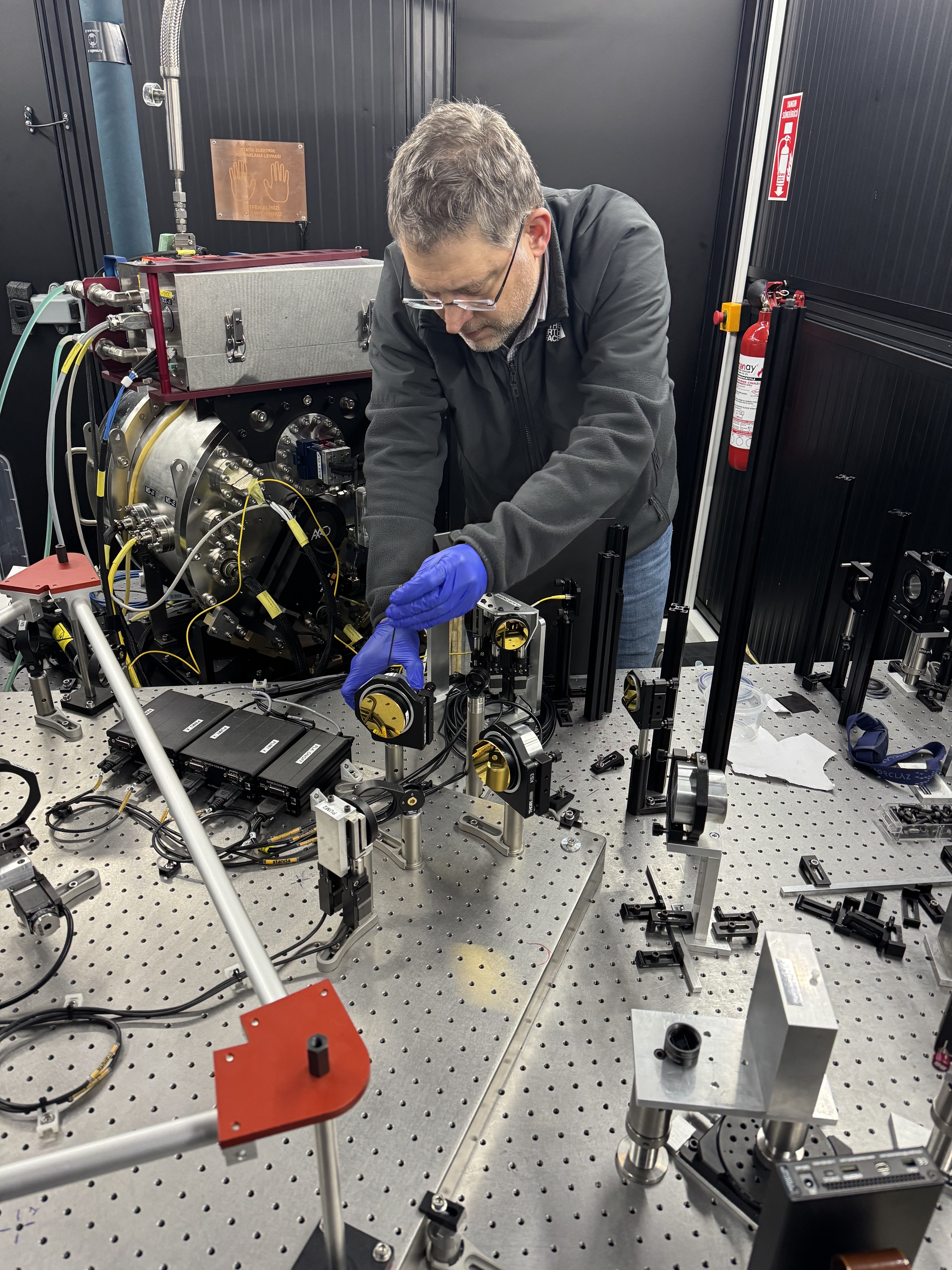}
    \end{tabular}
    \caption{PLACID alignment setup until reaching the SLM.}
    \label{fig:alignment1}
\end{figure}

After delivery to the DAG telescope, PLACID was installed on the diffraction-limited Nasmyth platform of the DAG telescope during the first semester of 2025, as well as cabled and successfully checked for functional integrity in mid-2025 \cite{2025SPIE13627E..0ZK}. Following the installation of the PLACID coronagraph on the Turkish 4-m DAG observatory in 2025, the on-site optical alignment using an internal light source and the preliminary acceptance (PA) measurements took place in early 2026. \\
The first step of the PLACID alignment campaign consisted of the exact placement of the coronagraph's breadboard, as well as the exact placement of the "PLACID on/off-stage" on the Nasmyth optical table, which either redirects light from TROIA to PLACID and back to DIRAC, or lets light pass from TROIA to DIRAC directly (see Figure \ref{fig:PLACIDplan}, the stage contains Pick-Up Mirror 1 (PUM1) and Put-Back Mirror 3 (PBM3)). \\
Once PLACID had been correctly placed with respect to the breadboard and consequently with respect to the other instruments, a monochromatic 633nm input beam was assembled from a fibered HeNe laser, a fiber launcher, a diaphragm, an off-axis parabolic mirror (OAP) and two fold mirrors (see top left image in Figure \ref{fig:alignment1}). This provision was required due to TROIA alignment work not yet having reached the point to produce an output beam. This input beam for PLACID therefore acts as a backup calibration source for PLACID alignment. For the beam to enter at the right height and position with respect to the table and the optics, the fine adjustment screws of the two fold mirrors that direct the beam towards the first fold mirror of the coronagraph, PUM1. This flat mirror is the first remotely actuated optical component, where the tip/tilt position can be adjusted via software, using the PLACID GUI - the others being PUM2, the Lyot stop wheel, PBM1, PBM2, and PBM3.

From there, an input pupil mask reproducing the DAG aperture (M1 with M2 secondary obstruction and support spiders) was placed in the beam, before reaching the second fold mirror PUM2 (see the Entrance Pupil Plane (EPP) in Figure \ref{fig:PLACIDplan}). The latter mirror directs the beam to the first parabolic mirror of PLACID (OAP1, f = 1000mm), which focuses the light onto the center of the SLM. Once mechanical fine tuning of PUM2, using the remotely controlled motors was achieved, a microscope objective with 20x magnification was installed to first inspect the PSF on a paper screen (see bottom left image in Figure \ref{fig:alignment1}). This procedure also entails adjusting the position of OAP1 accordingly, so the light would be centrally located on the spatial light modulator. The inspection of the PSF showed no obvious sign of low-order aberrations (astigmatism in particular) at this point. \\ 
At first, a "dummy SLM", which is just a flat mirror, was installed in the reflective coronagraphic focal-plane in place of the SLM and the beam reflected towards OAP2 (placed right next to OAP1, see bottom right image in Figure \ref{fig:alignment1} and plan in Figure \ref{fig:PLACIDplan}). 

\begin{figure}[h!]
    \centering
    \begin{tabular}{cc}
        \includegraphics[width=0.4\textwidth]{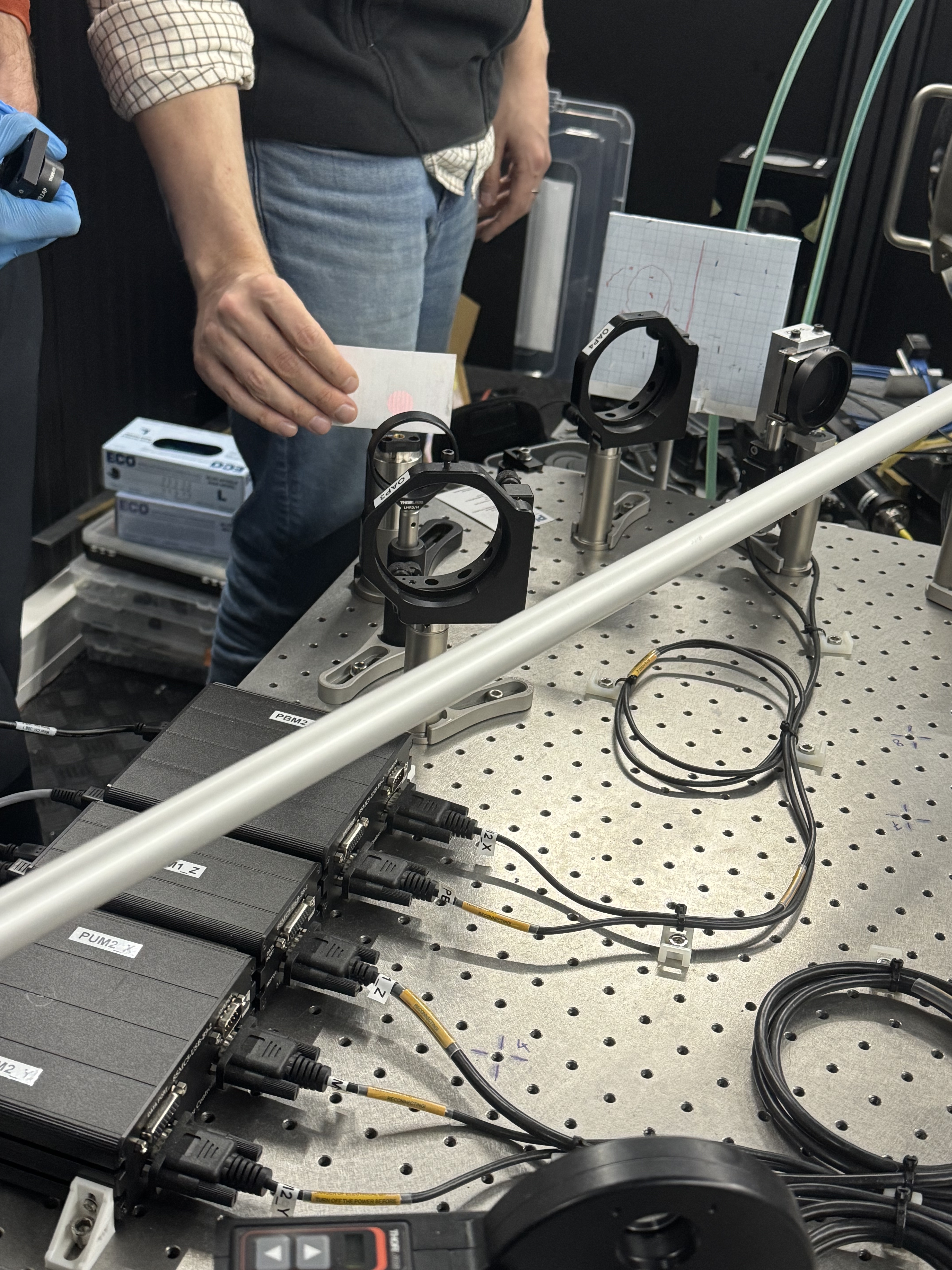} &
        \includegraphics[width=0.4\textwidth]{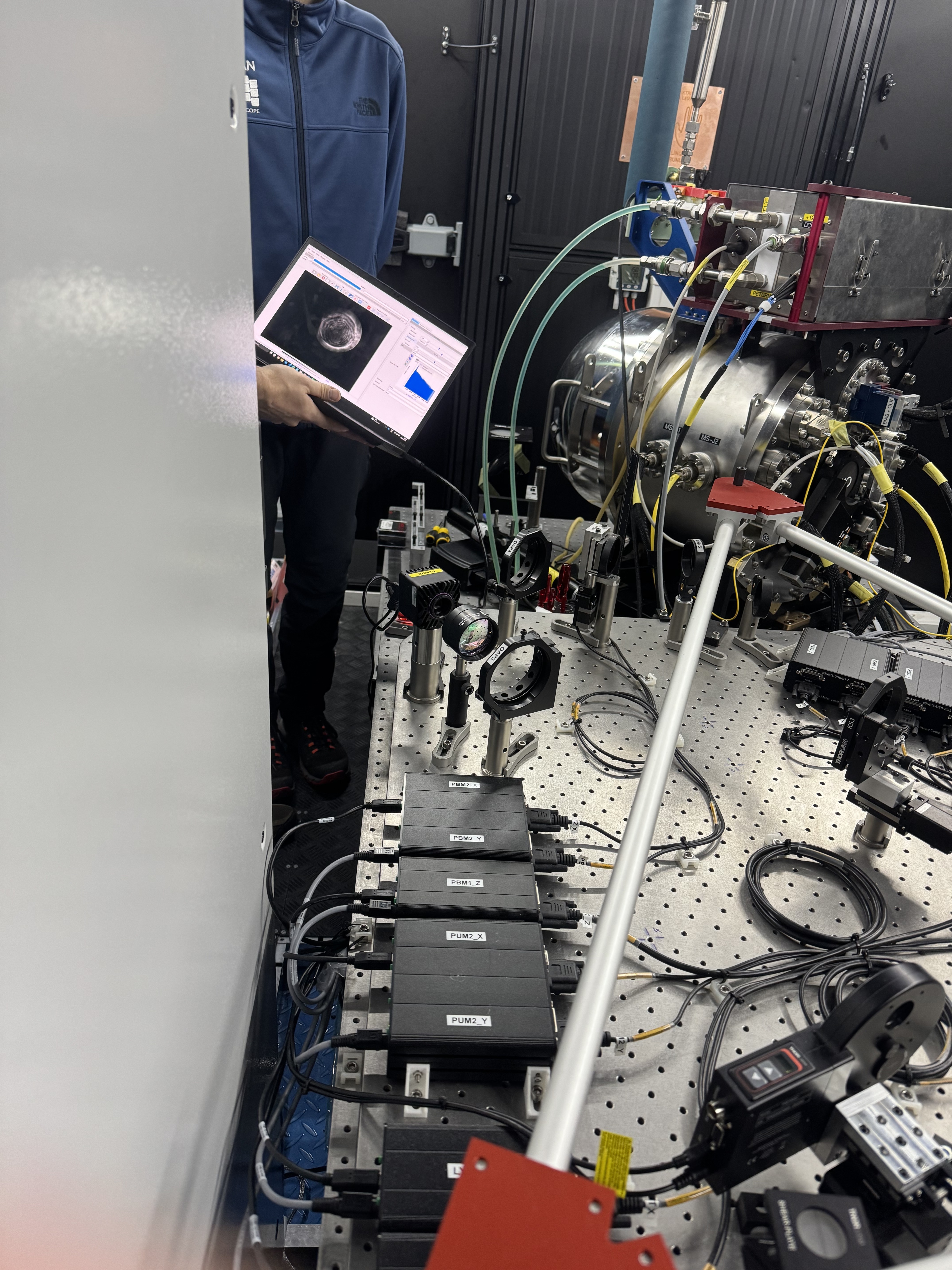} \\[0.5em]

        \multicolumn{2}{c}{
            \includegraphics[width=0.7\textwidth]{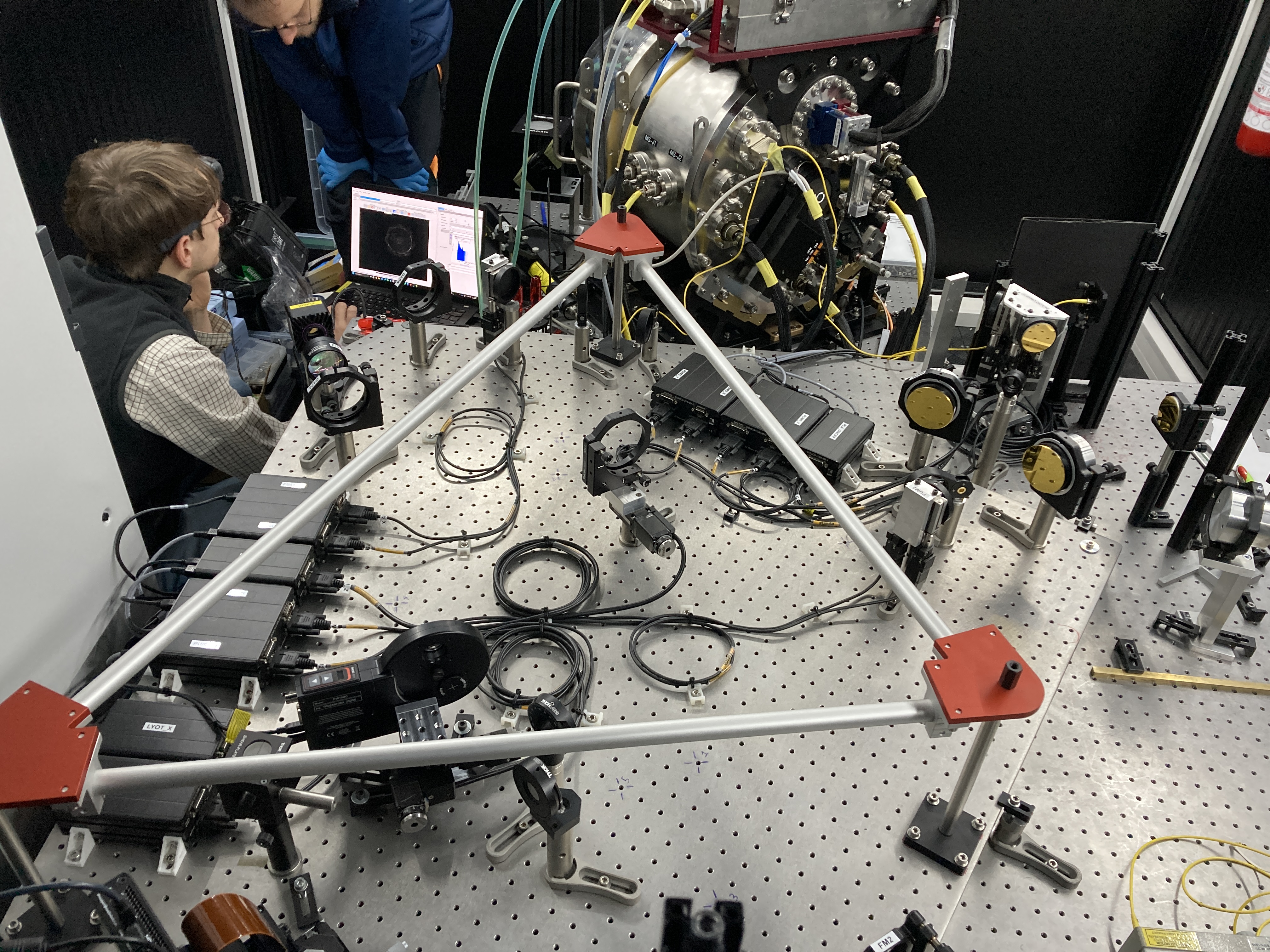}
        }
    \end{tabular}
    \caption[example]{\label{fig:alignment2}Inspection of the post-coronagraphic, post-Lyot pupil plane after alignment until the Lyot filter wheel.}
\end{figure}

For the time being, the flat mirror resided in the location of the SLM, until the divergent beam from the SLM dummy was successfully collimated by OAP2, reaches fold mirror 1 (FM1) and passes through the Lyot stop filter wheel. The collimation of the OAPs was always checked using a shear plate interferometer, which should show fringes parallel to the line of reference on its screen. \\
Soon after the beam position in the Lyot wheel proved successful (see top left image in Figure \ref{fig:alignment2}, with the Lyot wheel at the bottom of the image), the 1920 $\times$ 1152 pixel SLM from Meadowlark Corp. was installed in place of the dummy mirror in a coronagraphic reflective focal-plane (off-axis angle $\sim 5^{\circ}$). The modulator was subsequently activated and programmed with the default manufacturer flatmap. The position of the SLM in x-, y- and z-direction had to be fine tuned. The top right image in Figure \ref{fig:alignment1} shows the installation of the SLM in place. This was followed by the placement of a linear polarizer in front of the SLM, as the LCOS panel requires linearly polarized light to locally delay the path of light (phase shift). \\
Up until this point, the visible light source was used, however, as the Lyot pupil had to be imaged for alignment purposes (in particular to fine focus and center the SLM with respect to the beam), the light source was switched to a monochromatic 1550 nm source (Thorlabs S1FC1550 laser diode module) and a lens placed in the light path after the filter wheel to image the Lyot pupil onto a C-RED 3 InGaAs engineering camera (see top right and bottom image in Figure \ref{fig:alignment2}).

\begin{figure} [h!]
	\centering
   \includegraphics[width=0.5\textwidth]{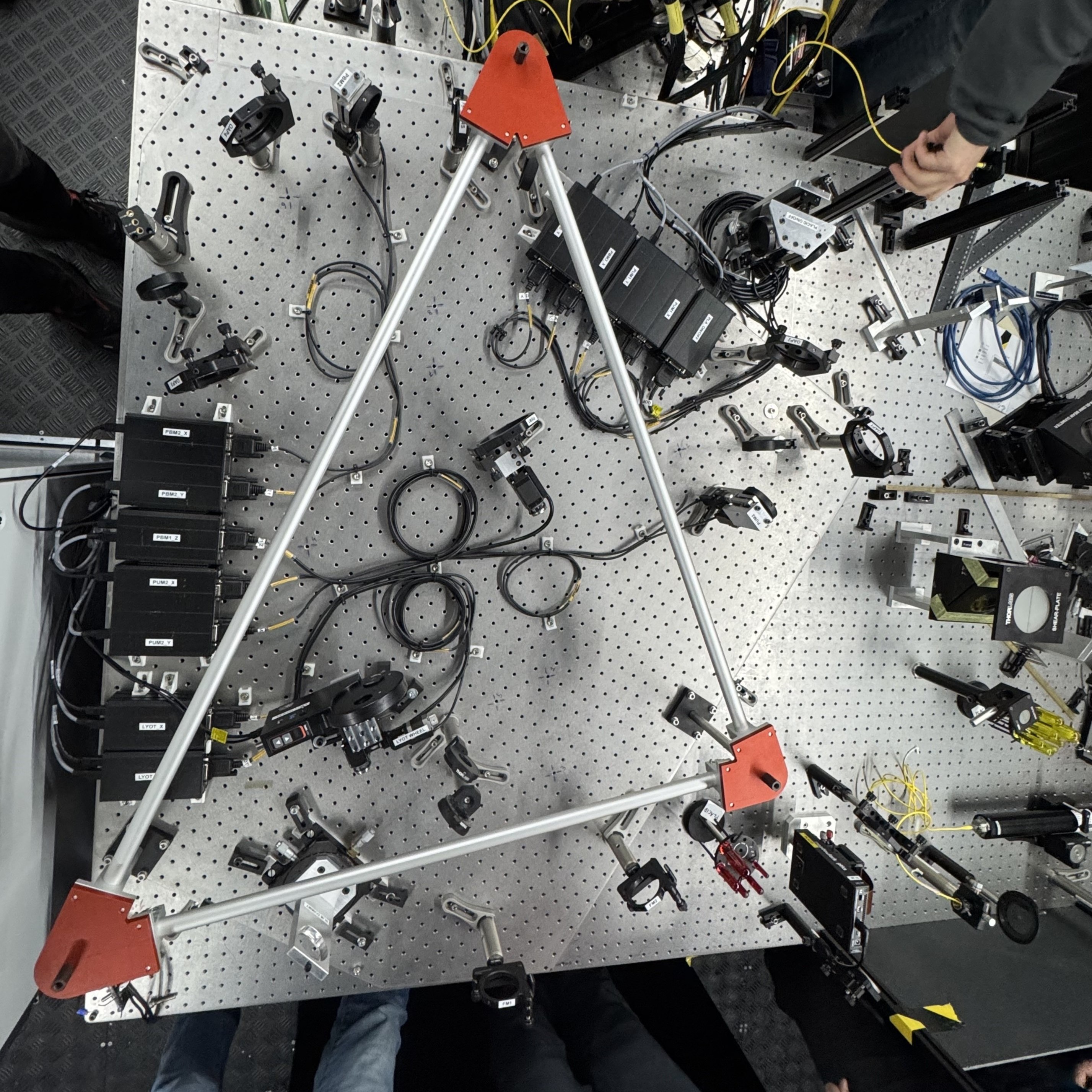} \\
   \includegraphics[width=0.5\textwidth]{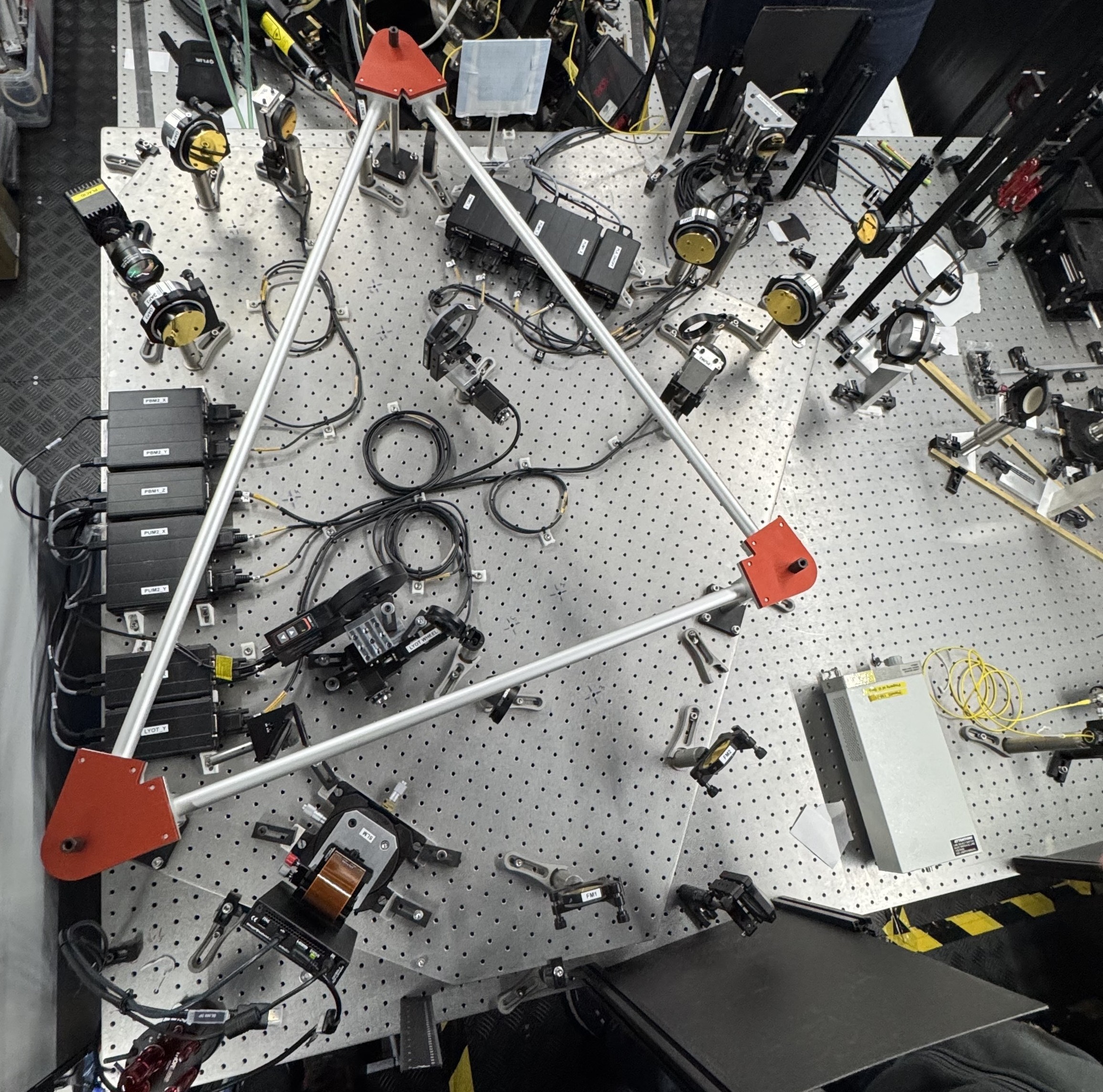}
   \caption[example] 
   { \label{fig:alignment3} Before and after (top to bottom) of the PLACID alignment campaign.}
\end{figure}

\begin{figure} [h!]
	\centering
   \includegraphics[width=0.8\textwidth]{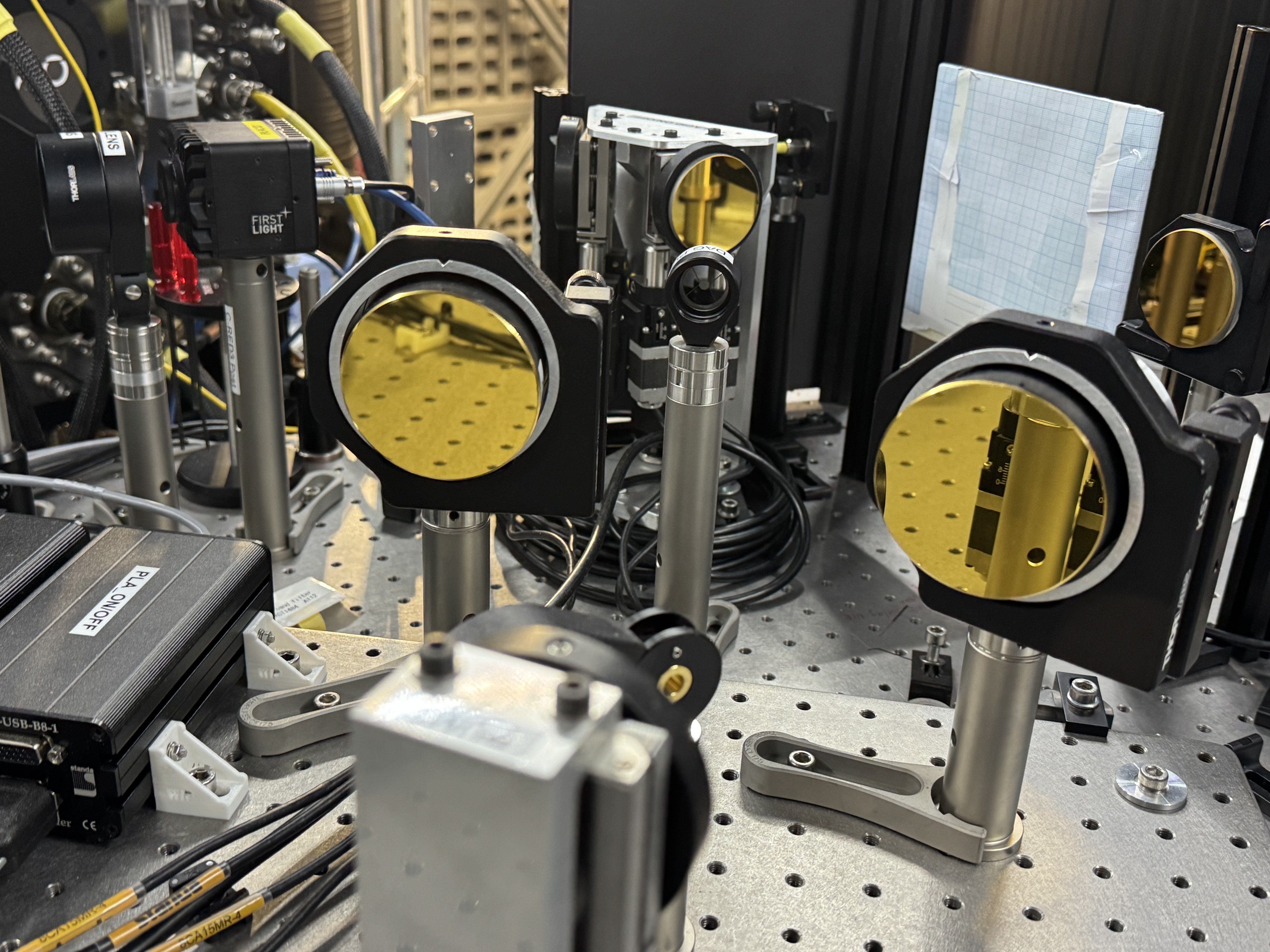} \\
   \includegraphics[width=0.8\textwidth]{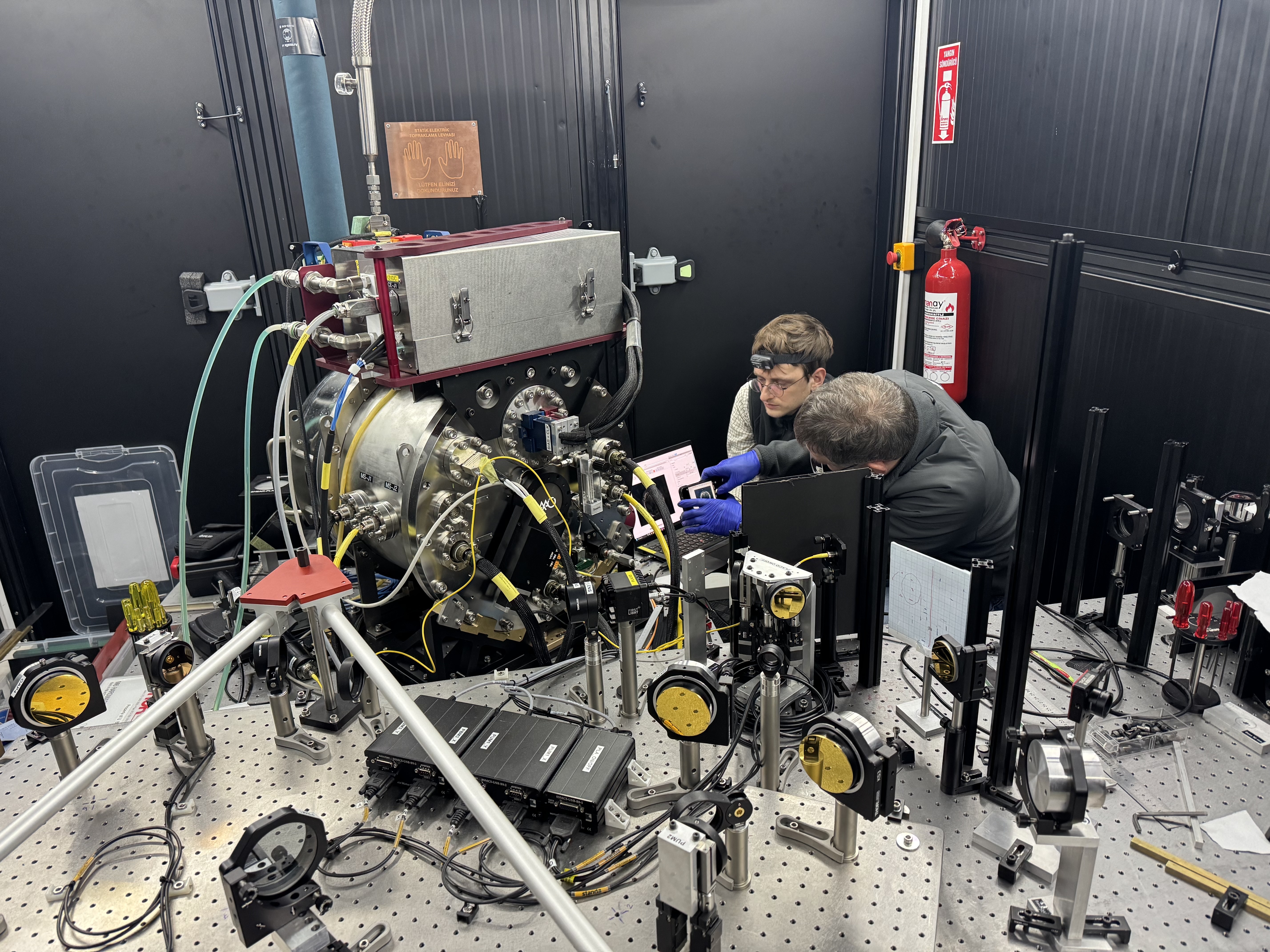}
   \caption[example] 
   { \label{fig:alignment4} Final alignment reached and first inspection of coronagraphic performance of PLACID.}
\end{figure}

Once the pupil plane tests proved successful, the light source was switched back to visible light, OAP3 placed inside its mount, and the central location of the beam on the optics confirmed. The focal point of OAP3's beam resides before FM2, making it possible to inspect the PSF once again with the microscope objective and to verify no significant aberrations had entered the system. We saw two PSFs caused by the SLM non-optimized A/R coating for visible light, and some speckles. Assessing the low-orders is hard at this point, but clear "cross-shapes" typical of astigmatism, or e.g. coma tails were not seen, so we could proceed. \\
This means that FM2 could easily be installed with the beam in its center, and soon thereafter, the placement of OAP4 provided a collimated beam towards the exit path of PLACID to be redirected step-by-step to DIRAC. At this point, the shear plate showed no noticeable astigmatism, which meant that the flat "put-back mirrors" PBM1 and PBM2 could easily be placed inside the path of light - again aligning the beam with respect to the mirror using the tip/tilt fine adjustment motors. PBM3 was not yet installed, as the DIRAC camera was not ready for use at this time. Instead, the aforementioned C-RED 3 infrared camera was installed between PBM2 and PBM3 (see top image in Figure \ref{fig:alignment4}). In Figure \ref{fig:alignment3}, the difference or before vs. after of the alignment of PLACID can be seen - the bottom image shows the optical mounts containing all the mirrors and the SLM positioned in place.

\begin{figure} [ht]
	\begin{center}
   		\begin{tabular}{cc} 
   		\includegraphics[width=0.45\textwidth]{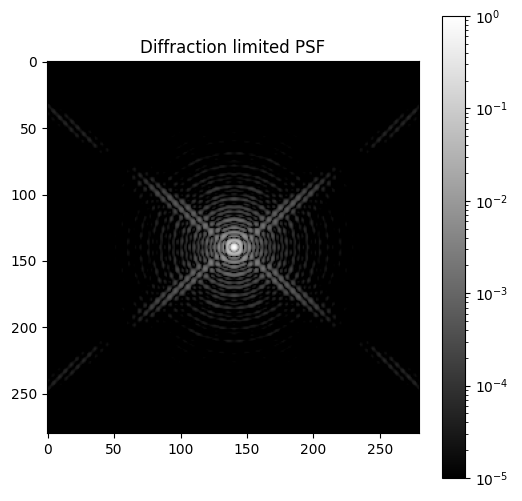} &
        \includegraphics[width=0.45\textwidth]{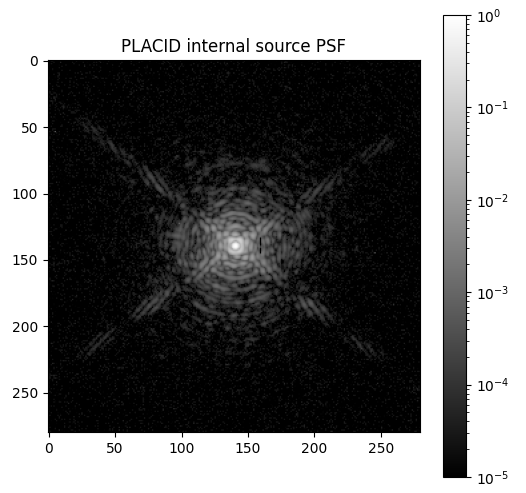} \\[0.5em]
		\end{tabular}
	\end{center}
   \caption[example] 
   { \label{fig:Diffraction_vs_Nasmyth}The theoretical diffraction limited PLACID PSF from simulation and the first measured PLACID PSF on-site for the monochromatic 1550 nm source.}
\end{figure}

With the successful alignment of the optics achieved, the next step was to inspect the PLACID output PSF on the C-RED 3 camera, first of all, using the monochromatic 1550 nm laser diode, and following suit, with an Ocean Insight Tungsten-Halogen white light (WL) source (OCOHL-2000-HP-FHSA) together with a spare DIRAC H-band filter place in front of the C-RED 3 camera. As can be seen in Figure \ref{fig:Diffraction_vs_Nasmyth}, the theoretical PLACID PSF from simulation (left image) and the measured first PSF using the monochromatic 1550 nm source (broadband), shows a very good match and the result can be called a success. There are more speckles than in the theoretical PSF, which is to be expected from the number of optical elements in the path.

\begin{figure} [h!]
	\centering
   \includegraphics[width=0.8\textwidth]{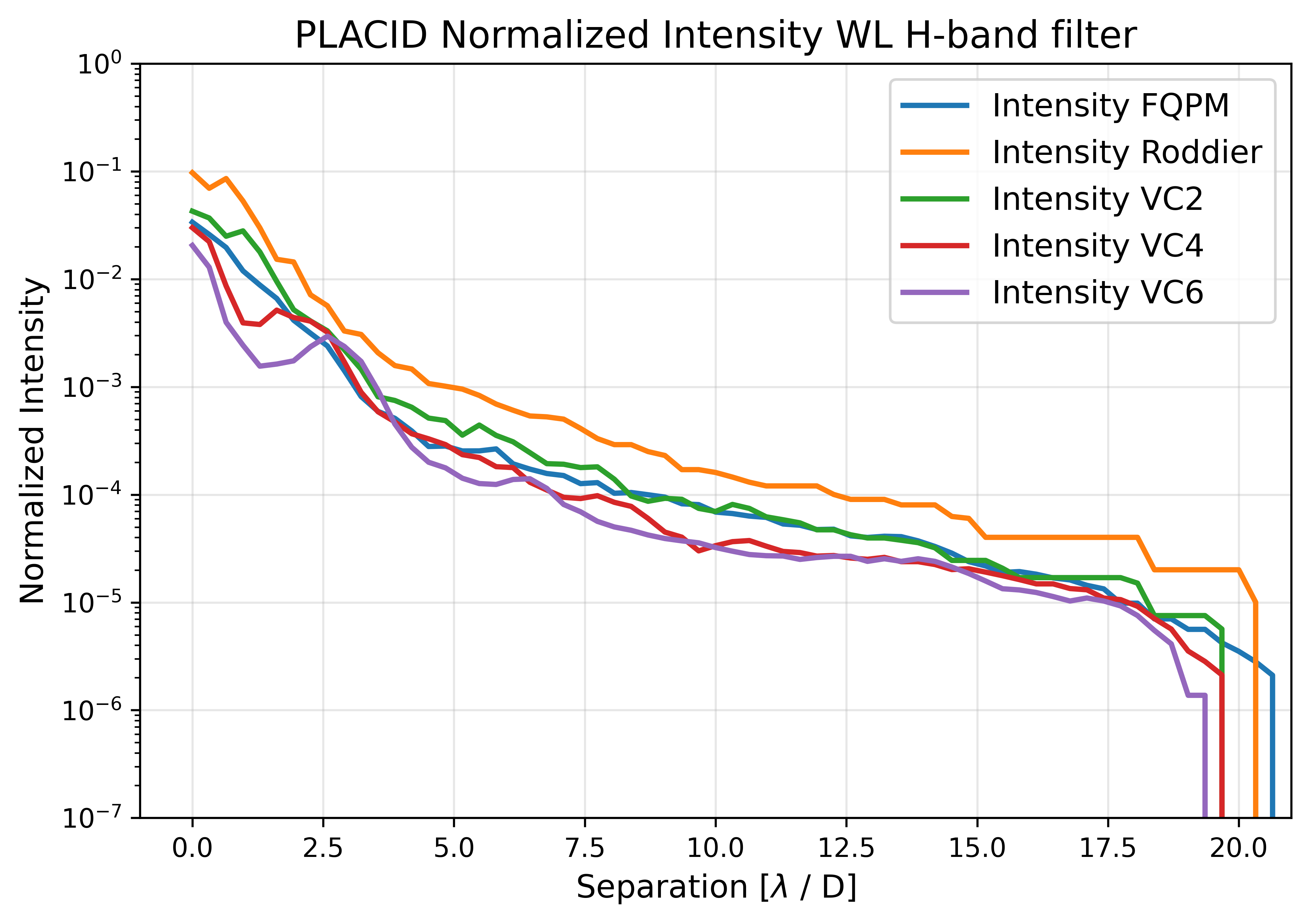}
   \caption[example] 
   { \label{fig:PA_intensity}Normalized intensity in function of angular separation for different coronagraphic masks applied on the SLM using a white light (WL) source and an H-band filter. Due to variable background conditions, the highest contrast does not exceed $10^{-5}$.}
\end{figure}

\begin{figure}[h!]
    \centering
        \includegraphics[width=0.87\textwidth]{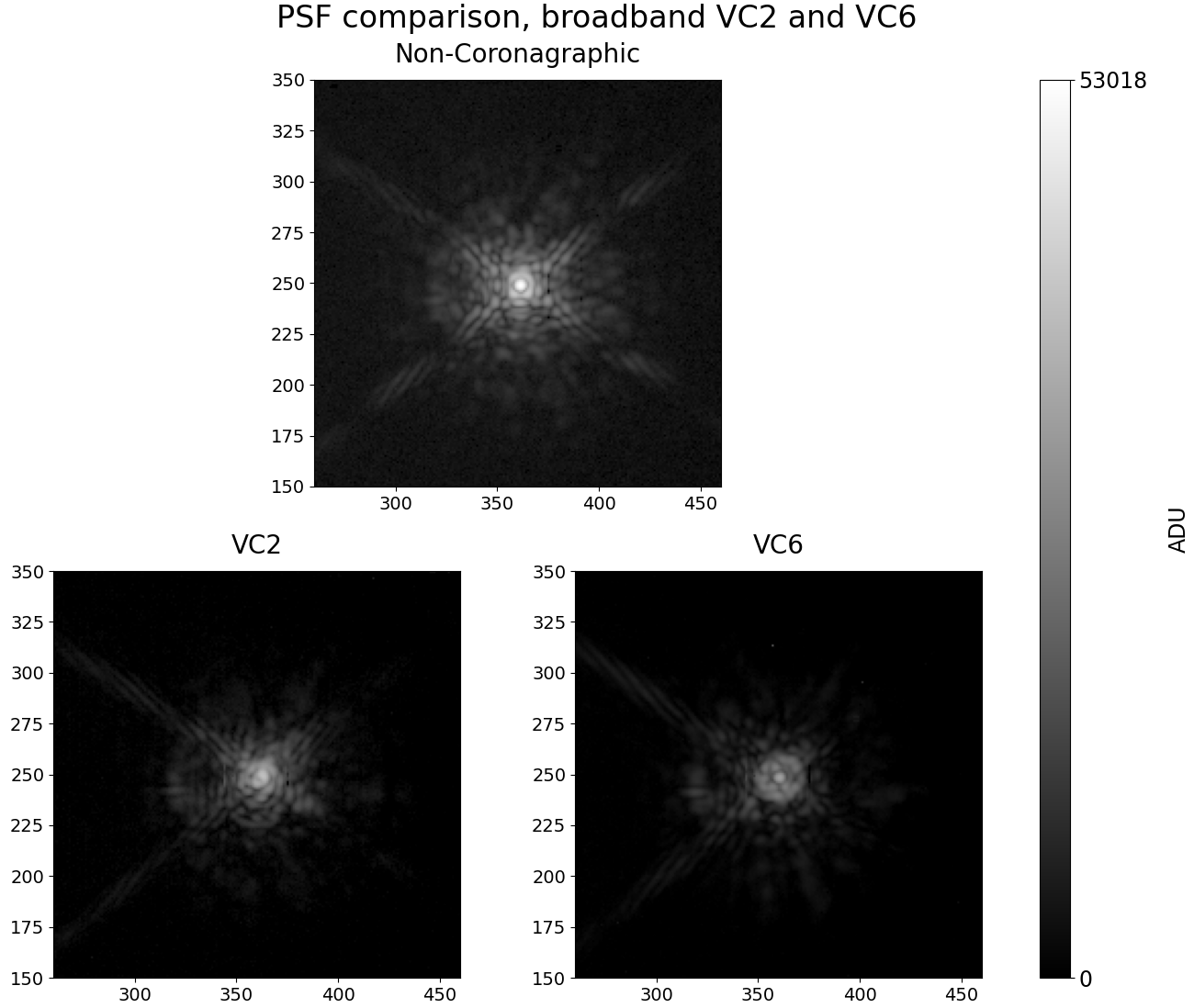}
    \caption[example]{\label{fig:PSFs}Non-coronagraphic PSF (top image) and coronagraphic PSFs for VC2 (bottom left) and for VC6 (bottom right) for broadband data acquisition.}
\end{figure}

The calibration and contrast performance of five different focal plane masks ensued, namely the Four Quadrant Phase Mask (FQPM, Rouan et al. 2000 \cite{2000PASP..112.1479R}), the Roddier\&Roddier mask (in short just "Roddier", Roddier\&Roddier 1997\cite{1997PASP..109..815R}), and the Vortex masks with topological charges 2, 4, and 6 (abbreviated as e.g. VC2, Foo et al. 2005 \cite{2005OptL...30.3308F}, Mawet et al. 2005 \cite{2005ApJ...633.1191M}, and Mawet et al. 2009 \cite{2009OExpr..17.1902M}). Furthermore, the PLACID GUI provides the option to tune a parameter named $\alpha$ that can adjust the slope of the look-up table translating 8-bits grey level into phase retardance. These values were fine tuned for each type of mask (FQPM, Roddier and the vortices) and thereafter applied to the mask for data acquisition. The data consisted of background frames (open-cap dark frames), coronagraphic, and non-coronagraphic PSFs (see also Figure \ref{fig:PSFs}). Subsequently, this data was reduced and the non-coronagraphic PSF center used as a reference to compute the radial separation when generating raw azimuthally averaged contrast curves, or more precisely "normalized residual intensity", not taking into account the respective coronagraphic throughput. \\
Something that was immediately evident during data acquisition with the C-RED 3 camera, was a residual pointing instability at short timescales when using the Roddier mask. The reasons for that are suspected to be the aforementioned temperature gradient on the Nasmyth platform (people present and moving, as well as the platform not being thermally enclosed long enough), causing ever so slight thermal deformation of the optics. The non-static state of the Nasmyth plaform also impacted background acquisition, causing the normalized intensity to flatten earlier than expected (see Figure \ref{fig:PA_intensity} and the outer contrast levels around $10^{-5}$). The instability of the Roddier mask for given conditions also lead to a re-evaluation of the approach to the binary star coronagraph mode for PLACID, which up until now solely utilized a double-Roddier phase mask (see more in section \ref{sec: binary_mode}).

\begin{figure}[h!]
    \centering
        \includegraphics[width=0.87\textwidth]{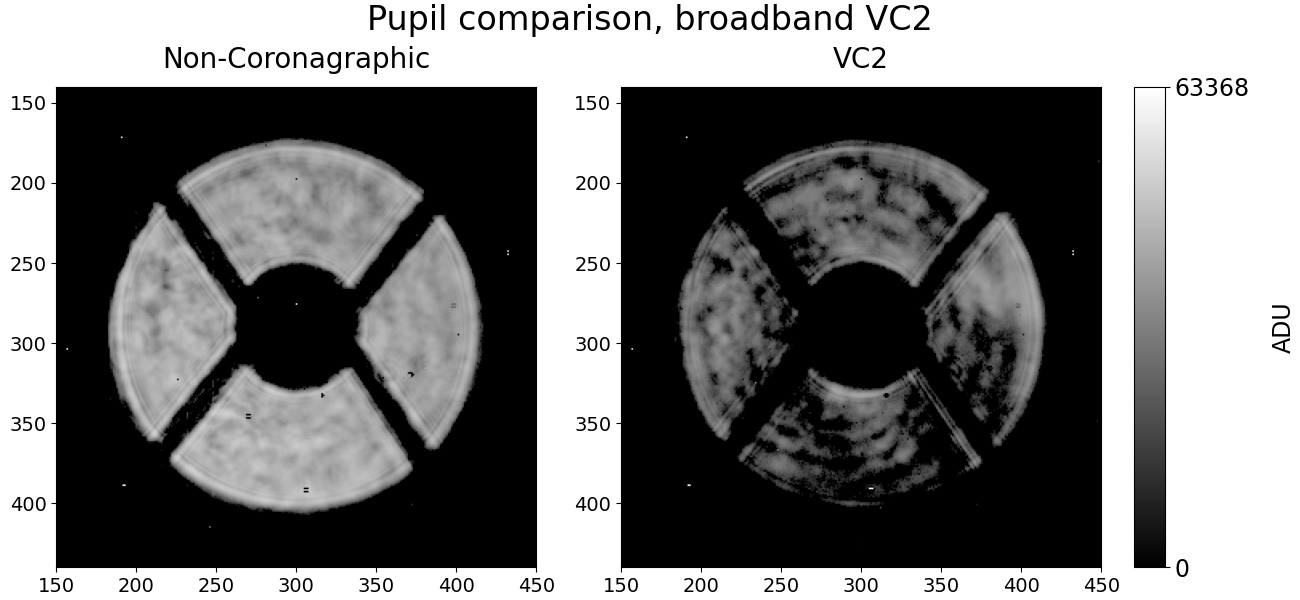}
    \caption[example]{\label{fig:Pupils}Non-coronagraphic pupil plane (left image) and the coronagraphic pupil for VC2 (right image) for broadband data acquisition.}
\end{figure}

As can be seen in Figure \ref{fig:PA_intensity}, the sequence of on-axis performance of the five coronagraphic masks is in ascending order: Roddier, VC2, FQPM, VC4, and VC6. The VC6 mask performs the best in terms of normalized intensity, but always comes with a large inner working angle (IWA) as opposed to the other masks tested here. FQPM and VC2 seem to perform similarly. Figure \ref{fig:PSFs} shows some images acquired with the C-RED 3 camera during preliminary acceptance, in focal plane configuration. The top image shows the PLACID PSF using the broadband white light source, the two bottom images show the coronagraphic PSFs for both VC2 and VC6, used to calculate the normalized intensity in Figure \ref{fig:PA_intensity}. Figure \ref{fig:Pupils} shows the pupil configuration without coronagraphic mask on the left and the pupil for a VC2 mask applied on the star to the right. 

\newpage

\section{PLACID DISCOVERY SPACE UPDATE} \label{sec: discovery_space}

The instrument’s on-sky discovery space had previously been estimated by using multiple databases, such as exoplanet.eu \footnote{\url{https://exoplanet.eu/home/}}, the NASA Exoplanet Archive \footnote{\url{https://exoplanetarchive.ipac.caltech.edu/}}, the Catalog of Circumstellar Disks \footnote{\url{https://www.circumstellardisks.org/}}, and the Washington Double Star catalog (WDS) \footnote{\url{https://www.astro.gsu.edu/wds/}}. These contain exoplanets, brown dwarfs, binaries, and disks and were added to the PLACID target list according to three exclusion criteria:

\begin{itemize}
\item Declination limit DEC $\geq -24^{\circ}$ (optimistic), due to the geographic location of the DAG telescope and considerations regarding airmass limitation (see Table \ref{tab:DAG})
\item Vmag $\leq 13$, due to the achievable Strehl ratio depending on brightness of the host star used as natural guide star (NGS) for the TROIA AO system  \cite{10.1117/12.3018836}
\item angular separation $\leq$ 16 arcsec, due to the PLACID FOV (see Table \ref{tab:PLACID})   
\end{itemize}

The previous iteration of the discovery space (presented in proceedings of SPIE \cite{10.1117/12.3018836}) included known directly imaged exoplanets and brown dwarfs in juxtaposition with raw lab contrast curves from factory acceptance, as well as a simple calculation of expected exposure time \cite{10.1117/12.3018836}.\\ 
In this work, we present an updated and improved version of the PLACID discovery space using a combination of preliminary acceptance (on-site) measurements and simulations. These parameters are fed into an updated exposure time calculator for PLACID, calculating the collected photons for star and planet, and accounting for 

\begin{itemize}
\item Telescope collecting area
\item H-band filter bandpass
\item Contrast at given angular separation, consisting of measured normalized intensity and measured throughput for the applied VC2 mask at preliminary acceptance (February 2026)
\item Signal-to-noise-ratio (SNR) equal to 5
\item Instrument transmission of 20\% (estimation based on factory acceptance data)
\item Strehl ratio of 75\% (pessimistic assumption)
\item Sky background
\item Dark current
\item Read-out noise
\item Photometric aperture size.
\end{itemize}

\begin{figure} [h!]
	\begin{center}
   		\begin{tabular}{c} 
   		\includegraphics[width=0.8\textwidth]{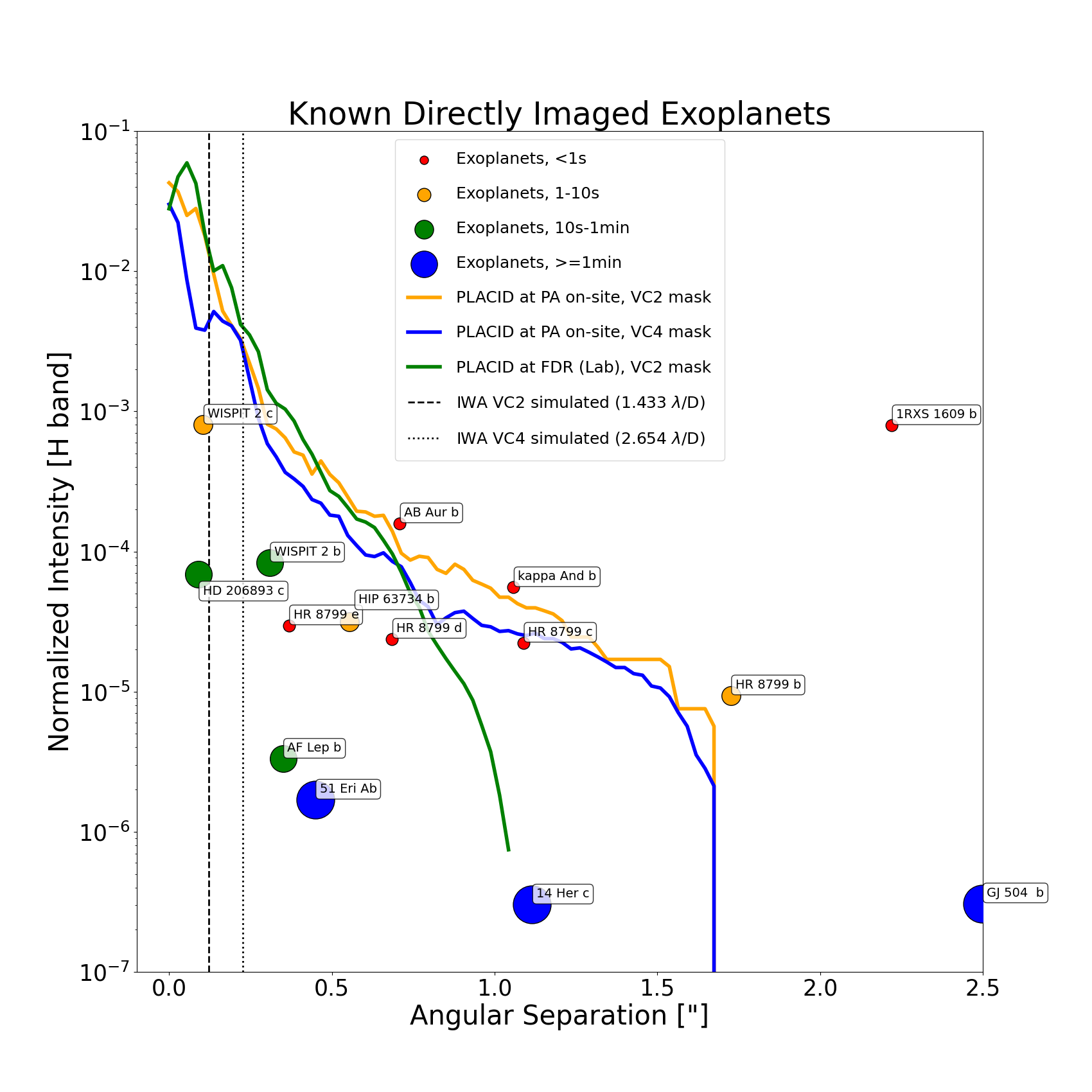}
		\end{tabular}
	\end{center}
   \caption[example] 
   { \label{fig:Exoplanets}Updated exoplanet discovery space for the PLACID coronagraph, using PLACID factory acceptance (in the lab, FDR) and preliminary acceptance tests (PA) on-site as a reference, and showing expected exposure time and Inner Working Angle (IWA) for two different masks.}
\end{figure}

\begin{figure} [h!]
	\begin{center}
   		\begin{tabular}{c} 
   		\includegraphics[width=0.8\textwidth]{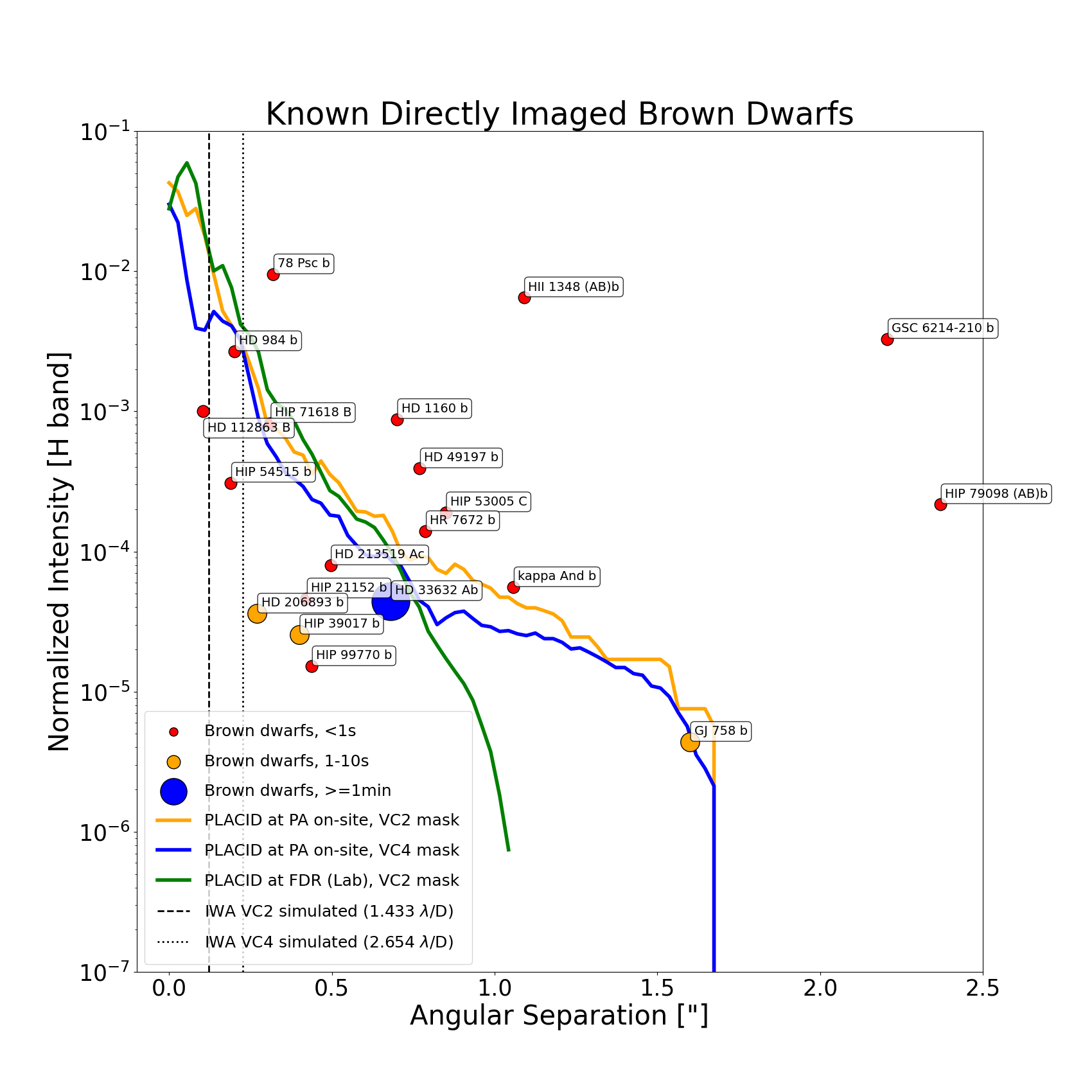}
		\end{tabular}
	\end{center}
   \caption[example] 
   { \label{fig:Brown_Dwarfs}Updated brown dwarf discovery space for the PLACID coronagraph, using PLACID factory acceptance (in the lab, FDR) and preliminary acceptance tests (PA) on-site as a reference, and showing expected exposure time and Inner Working Angle (IWA) for two different masks.}
\end{figure}

However, this model does not include a PSF model, saturation model, ADI self-subtraction, field rotation, atmospheric variability, detector non-linearity, flat-field errors, imperfect background subtraction, or temporal evolution of speckles. The calculated exposure time therefore assumes that residual stellar speckles behave like ordinary photon noise and can be averaged down indefinitely with longer exposure time, which is why the values as shown in Figures \ref{fig:Exoplanets} and \ref{fig:Brown_Dwarfs} should be taken with a grain of salt.\\ 
This information should serve as an estimate how difficult or easy it could be to observe a given target and is meant to support local and visiting astronomers in planning early high-contrast imaging programs, as well as planning the first commissioning and possibly science observations for PLACID in the coming months.

\begin{table}[h!]
\caption{PLACID specifications} 
\label{tab:PLACID}
\begin{center}       
\begin{tabular}{|l|l|} 
\hline
\rule[-1ex]{0pt}{3.5ex}  \textbf{Name} & \textbf{PLACID}  \\
\hline
\rule[-1ex]{0pt}{3.5ex}  \textbf{Observing bands} & H-band: 1.65 $\mu$m (Ks-band: 2.15 $\mu$m)   \\
\hline
\rule[-1ex]{0pt}{3.5ex}  \textbf{SLM specification} & 1920 x 1152 px, 8 bits  \\
\hline
\rule[-1ex]{0pt}{3.5ex}  \textbf{SLM spatial resolution} & 10 px/$\frac{\lambda}{D}$  \\
\hline
\rule[-1ex]{0pt}{3.5ex}  \textbf{FOV} & 16" x 9.6"  \\
\hline
\rule[-1ex]{0pt}{3.5ex}  \boldmath$\lambda/D$ \textbf{at H-band} & 85.084 mas \\
\hline
\rule[-1ex]{0pt}{3.5ex}  \textbf{Optical throughput} & $> 20$\%  \\
\hline
\end{tabular}
\end{center}
\end{table}

At first glance, Figures \ref{fig:Exoplanets} and \ref{fig:Brown_Dwarfs} clearly display, that some known directly imaged exoplanets and brown dwarfs will pose a challenge to observe, while others seem feasible. Specifically, \textit{kappa And b} (often also categorized as a brown dwarf, see more recently Currie et al. 2018 \cite{2018AJ....156..291C} and Godoy et al. 2025 \cite{2025A&A...702A...4G}, also plotted in Figure \ref{fig:Brown_Dwarfs}), \textit{HR 8799 b} and \textit{c}, \textit{AB Aur b} and \textit{1RXS 1609 b} appear to be in the realm of achievable detections, while \textit{HR 8799 d} and \textit{e}, and \textit{WISPIT 2 b} for example, will be limiting cases, that could be detected after post-processing algorithms such as ADI are implemented (as is the case in the PLACID data reduction software). With post-processing techniques, we hope to achieve improvements in contrast around a factor of 10. Other targets on this graph will likely remain undetected, like \textit{AF Lep b} and \textit{51 Eri Ab}. The discrepancy between the factory and preliminary acceptance normalized intensity from a separation of 1" onwards can likely be traced back to the instable conditions in the Nasmyth platform during the last campaign, where background fluctuations were high, as was mentioned earlier.

Similarly, for the brown dwarfs in Figure \ref{fig:Brown_Dwarfs}, there are a significant number of targets above the plotted curves, indicating they will not pose a challenge when attempting to observe them. In fact, comparing the brown dwarf to the exoplanet graph, the brown dwarfs tend to be slightly brighter as compared to the exoplanets, which would make observing them easier.

\begin{figure} [h!]
	\begin{center}
   		\begin{tabular}{c} 
   		\includegraphics[width=0.95\textwidth]{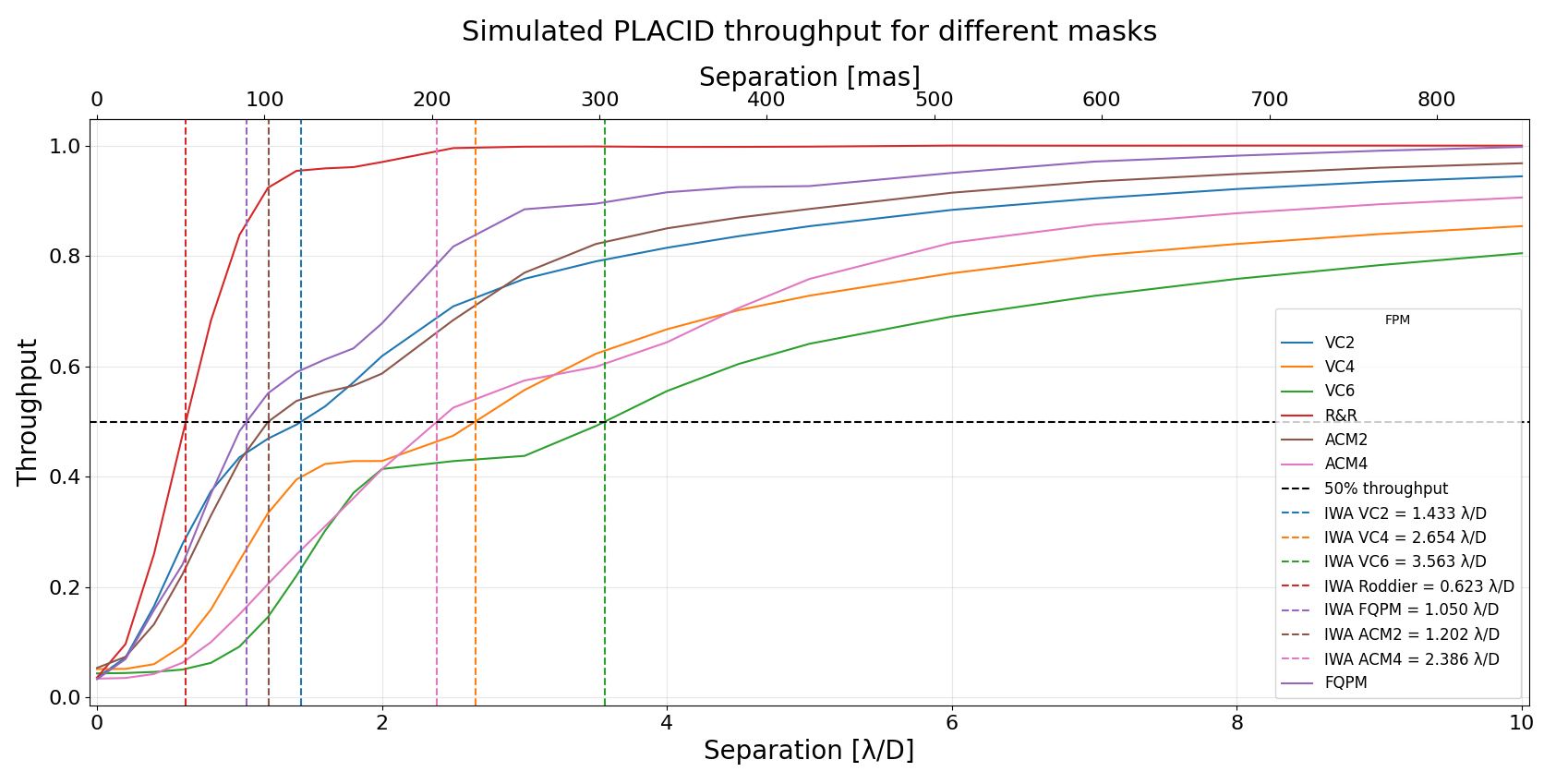}
		\end{tabular}
	\end{center}
   \caption{\label{fig:throughput_sim}Simulated broadband PLACID throughput (with telescope pupil and Lyot stop) for seven different phase masks: vortex masks of charges 2, 4 and 6, Roddier, FQPM, and Azimuthal Cosine Masks (ACM) of charges 2 and 4 for H-band.}
\end{figure}

Finally, two vertical lines were plotted, that represent the IWA for a VC2 and VC4 mask, respectively. These values arise from simulations assuming throughput as a function of angular separation for the simulated focal-plane masks, while using an 8-bit greyscale implementation (required for the SLM), a central obstruction with spiders (secondary mirror with mount) in the entrance pupil, a sampling of 10 pixels per $\lambda/D$ and setting the Lyot stop to be the same as the entrance pupil. This simulation was applied to a set of masks, namely the VC2, VC4, and VC6, Roddier, FQPM, but also the Azimuthal Cosine Mask (ACM, Hénault et al. 2015 \cite{2015JOSAA..32.1276H}, Hénault et al. 2018 \cite{2018OptCo.423..186H}) of charge 2 and 4. The results for that can be seen in Figure \ref{fig:throughput_sim} and in Table \ref{tab:IWA}. All masks are evaluated along the horizontal direction except FQPM, which is evaluated along the diagonal direction, as the 0 and $\pi$ demarcation line proceeds in a horizontal direction.  

As can clearly be seen, the Roddier\&Roddier phase mask possesses a very good ratio of central suppression of the star and quick recovery toward a high throughput. According to the on-site measurements at preliminary acceptance, this may however look a bit different when accounting for sensitivity to tip/tilt variations. The FQPM's sensitivity to jitter of that kind may also deteriorate its performance once the instrument is on-sky. The ACM masks perform consistently better (at least in terms of IWA - for on-axis nulling, ACM2 performs worse than VC2) than their vortex counterparts, which can be clearly seen in both Table \ref{tab:IWA} and Figure \ref{fig:throughput_sim} and could be an interesting avenue to explore once commissioning activities commence. The ACM differs from the vortex mask such that instead of using the full phase ramp from 0 to $2 \pi$, it goes from 0 to 2 $\times$ the first zero of the Bessel function $J_0$ ($b_0 = 2.4048$ rad and $2 b_0 = 4.8096$ rad \cite{2018OptCo.423..186H}).

\begin{table}[h!]
\caption{\label{tab:IWA}Inner Working Angle (IWA) for vortex masks of charges 2, 4 and 6, Roddier, FQPM, and Azimuthal Cosine Masks (ACM) of charges 2 and 4 for H-band.}
\begin{center}
\begin{tabular}{|c|c|c|}
\hline
\rule[-1ex]{0pt}{3.5ex}  & \textbf{IWA {[}$\lambda$/D{]}} & \textbf{IWA {[}mas{]}}  \\ \hline
\rule[-1ex]{0pt}{3.5ex}  \textbf{VC2} & 1.433                 & 121.915  \\ \hline
\rule[-1ex]{0pt}{3.5ex} \textbf{VC4} & 2.654                 & 225.820 \\ \hline
\rule[-1ex]{0pt}{3.5ex} \textbf{VC6} & 3.563                 & 303.178  \\ \hline
\rule[-1ex]{0pt}{3.5ex} \textbf{Roddier} & 0.623              & 53.048  \\ \hline
\rule[-1ex]{0pt}{3.5ex} \textbf{FQPM} & 1.050                 & 89.355  \\ \hline
\rule[-1ex]{0pt}{3.5ex} \textbf{ACM2} & 1.202                & 102.283  \\ \hline
\rule[-1ex]{0pt}{3.5ex} \textbf{ACM4} & 2.386                 & 203.053  \\ \hline
\end{tabular}
\end{center}
\end{table}

\newpage

\section{THE PLACID BINARY STAR MODE} \label{sec: binary_mode}

\begin{figure}[htbp]
    \centering
        \includegraphics[width=0.8\textwidth]{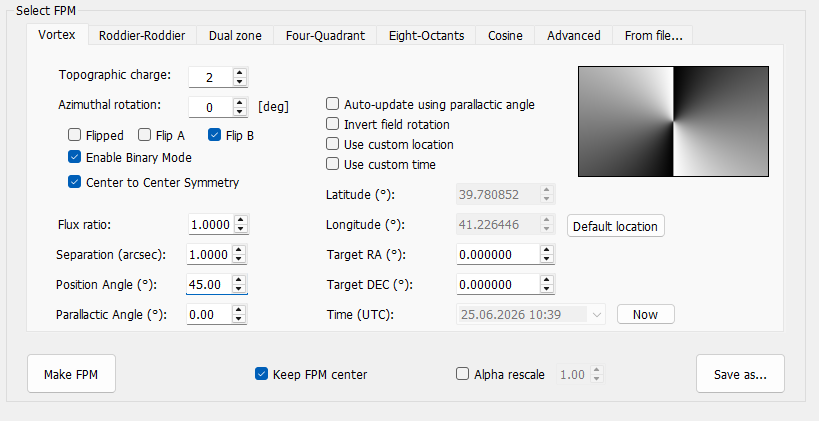}
    \caption{The PLACID GUI FPM section, with the binary mode enabled for a vortex mask.}
    \label{fig:binary_GUI}
\end{figure}

\begin{figure}[htbp]
    \centering
        \includegraphics[width=0.6\textwidth]{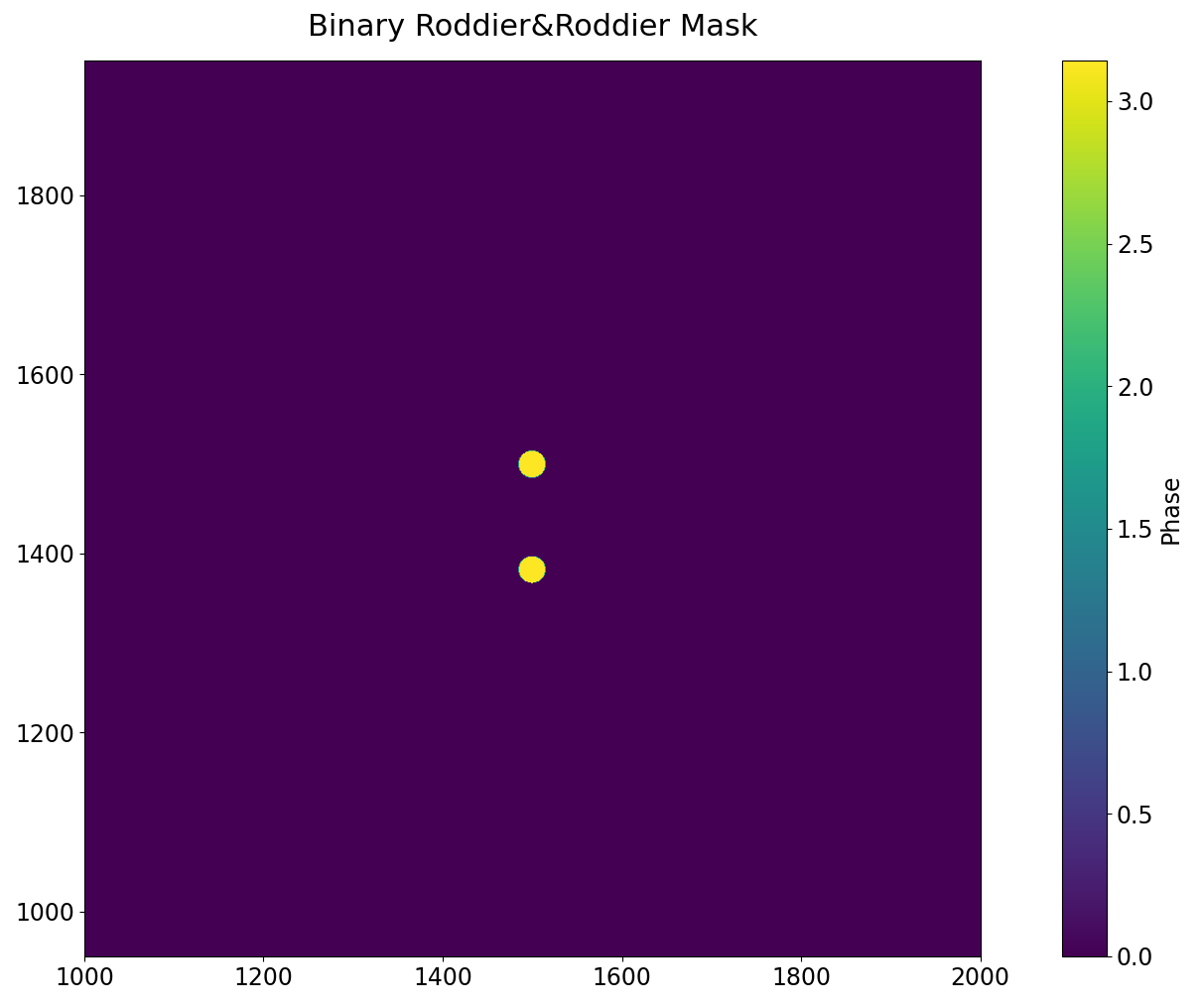}
    \caption{The double Roddier mask for an SLM with 10 px per $\lambda$/D. The Roddier dots have a radius of 15 px to make them easily visible.}
    \label{fig:Roddier_binary}
\end{figure}

As mentioned before, the SLM in the focal plane of PLACID can not only adapt to observing conditions but also to specific targets. The fact that any imaginable phase pattern can be programmed onto the modulator, leads to the logical conclusion, that double or even multiple versions of commonly used FPMs can be applied. Current estimates regarding the abundance of binary systems in our galaxy go towards 50\% or more \cite{2010ApJS..190....1R} - this means, that PLACID could potentially open the door to observing many more of the planetary systems that were not possible in direct imaging up until this point. Applying direct imaging to binary systems already exists in different forms (e.g. Cady et al. 2011 \cite{2011PASP..123..333C}) - one of which is placing two focal plane masks in cascade (see K\"uhn et al. 2018 \cite{2018SPIE10702E..42K}). These binary direct imaging configurations can be considered complicated and uncompromising - the SLM-approach in the focal plane simplifies things significantly. With only one mask required, no need for manufacturing, and angular separations as well as contrast ratios between primary and secondary component being adjustable with just a quick change in the controls, PLACID seems ideal for the task. \\
The first implementation of this concept on the PLACID GUI is also the simplest - using a double Roddier coronagraph, which consists of just two dots of $\pi$ on a background of zero phase shift (see Figure \ref{fig:Roddier_binary}). The distance between the two points, and the position angle between the two components of the binary are input parameters of the GUI for generating the mask (see Figure \ref{fig:binary_GUI}). Another advantage of using an SLM for this observing mode, is the implementation of binary ADI. This means that the relative motion caused by field rotation of the sky for altitude/azimuth telescopes can be used to the advantage of achieving higher contrast levels. We fix the primary component of the binary with one mask, while the other mask center of the secondary moves around the primary during the observation run. This PSF subtraction would, however, only work for the primary component, on which the observation is centered. Subtracting both stellar PSFs in a binary would require derotation, which would strip ADI of its meaning, as it requires rotation in order not to self-subtract the planet. Therefore, either an approach of combining both ADI and Reference Differential Imaging (RDI) or just applying RDI alone for any stellar component would be required.

In the automatically updating binary ADI-mode of the GUI, the algorithm can adapt for the relative position of the binary components by calculating the parallactic angle, using angular separation, position angle, the coordinates of the target (in right ascension (RA J2000) and declination (DEC J2000)), the location of the observation on Earth, and the observation time as input parameters. 

\begin{figure}[htbp]
    \centering
    \begin{tabular}{cc}
        \includegraphics[width=0.45\textwidth]{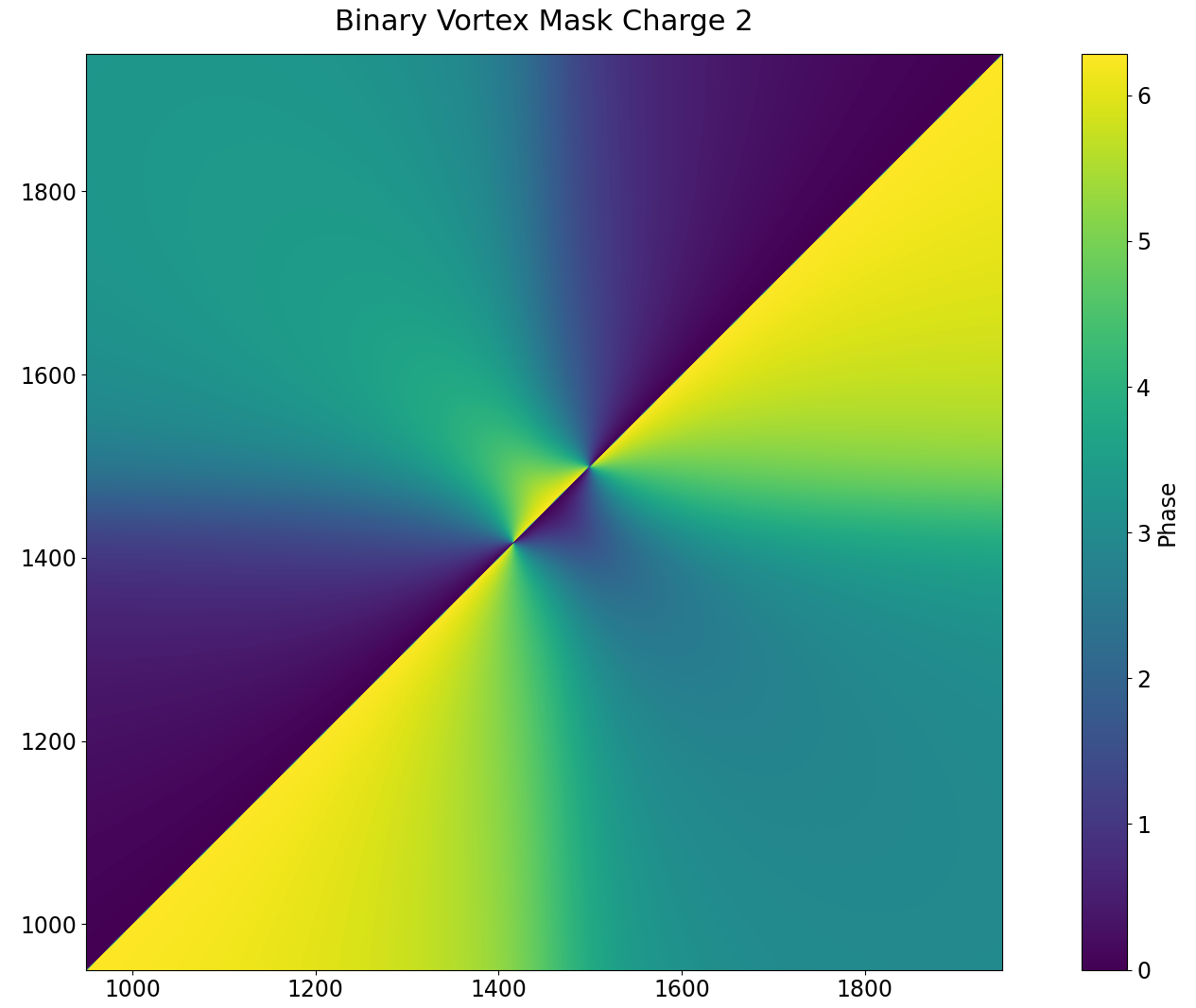} &
        \includegraphics[width=0.45\textwidth]{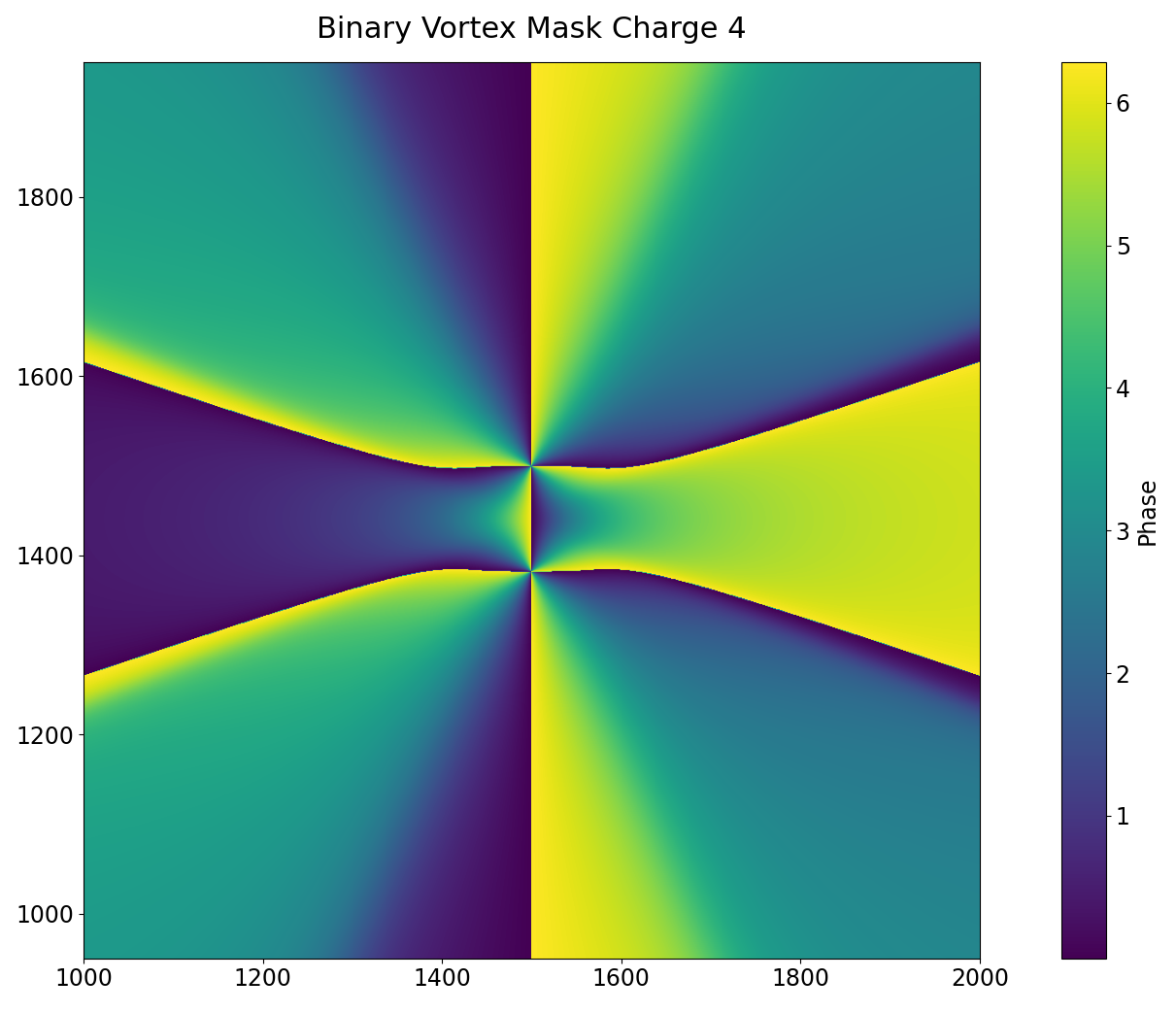} \\[0.5em]
        \includegraphics[width=0.45\textwidth]{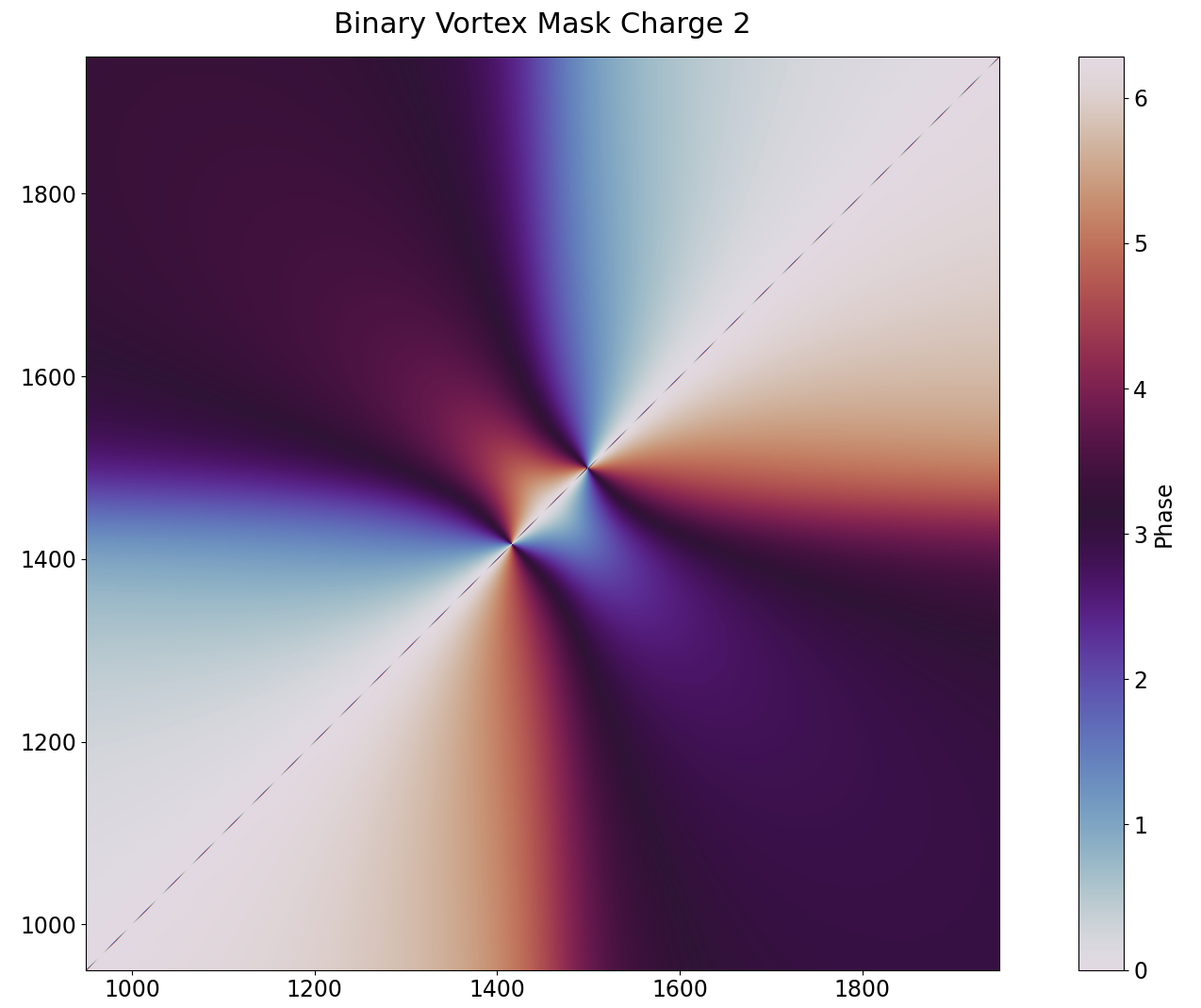} &
        \includegraphics[width=0.45\textwidth]{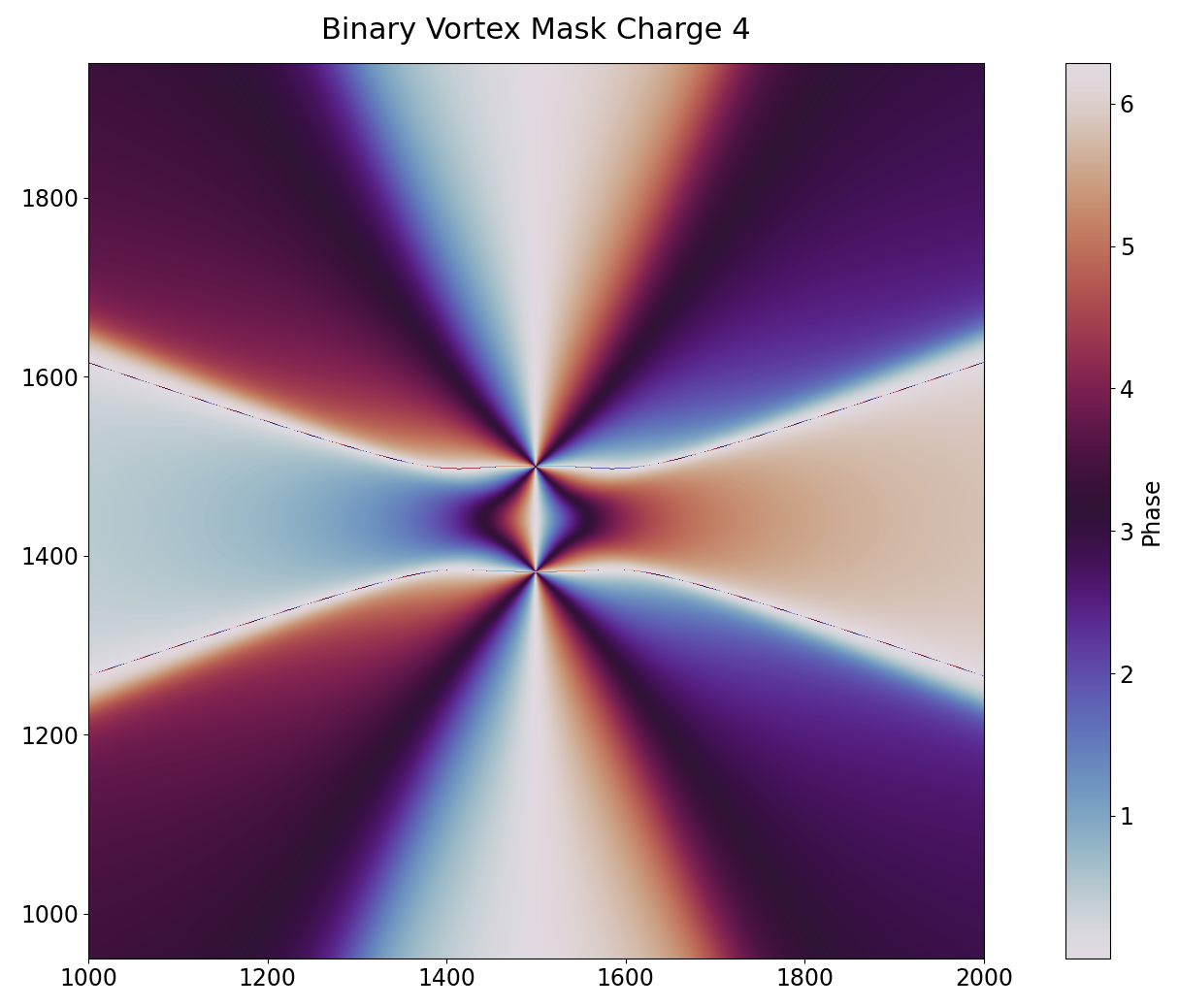} \\[0.5em]
    \end{tabular}
    \caption{Binary VC2 and VC4 masks with secondary component flipped relative to the primary. The top row images represent the phase shifts as implemented on the SLM, the bottom row images the phase shift applied physically to the light, using a color map that is white for 0 and $2 \pi$.}
    \label{fig:VC2_VC4_binary}
\end{figure}

Due to the lack of robustness of the Roddier mask with respect to residual tip/tilt jitter, as already observed during the last February run, the approach required alternate pathways, namely, by creating new binary masks, in particular binary vortices and ACMs. Two examples of that are the binary VC2 and VC4 masks as shown in Figure \ref{fig:VC2_VC4_binary}. The top row images in Figure \ref{fig:VC2_VC4_binary} show how the respective masks would be represented on the SLM pixels, being within the range of $0$ to $2 \pi$. The borders between those two values (where yellow and blue meet) seem like hard phase offsets, but are in fact smooth transitions, as a phase shift of $0$ or $2 \pi$ is physically equivalent. This is why the bottom images in both figures  show more physically accurate phase transitions, as the color map starts and ends in white. The difficulty of generating these masks lies in the merging of two vortex functions - in this case, two weighting functions were applied to give the two masks more or less "dominance" within a region of the mask center. This influence or weighting function also makes it possible to treat binaries with different magnitudes accordingly - the input parameter for contrast ratio in Figure \ref{fig:binary_GUI} is a factor that gives priority for the primary component over the secondary if the ratio is smaller than 1. This approach can be applied to detect circumbinary (P-type) disks or companions, but of course also for planets around one component (S-Type) for wider separation systems.

\newpage

\section{DISCUSSION AND OUTLOOK} \label{sec:discussion}

\begin{figure} [ht]
	\begin{center}
   		\begin{tabular}{c} 
   		\includegraphics[width=\textwidth]{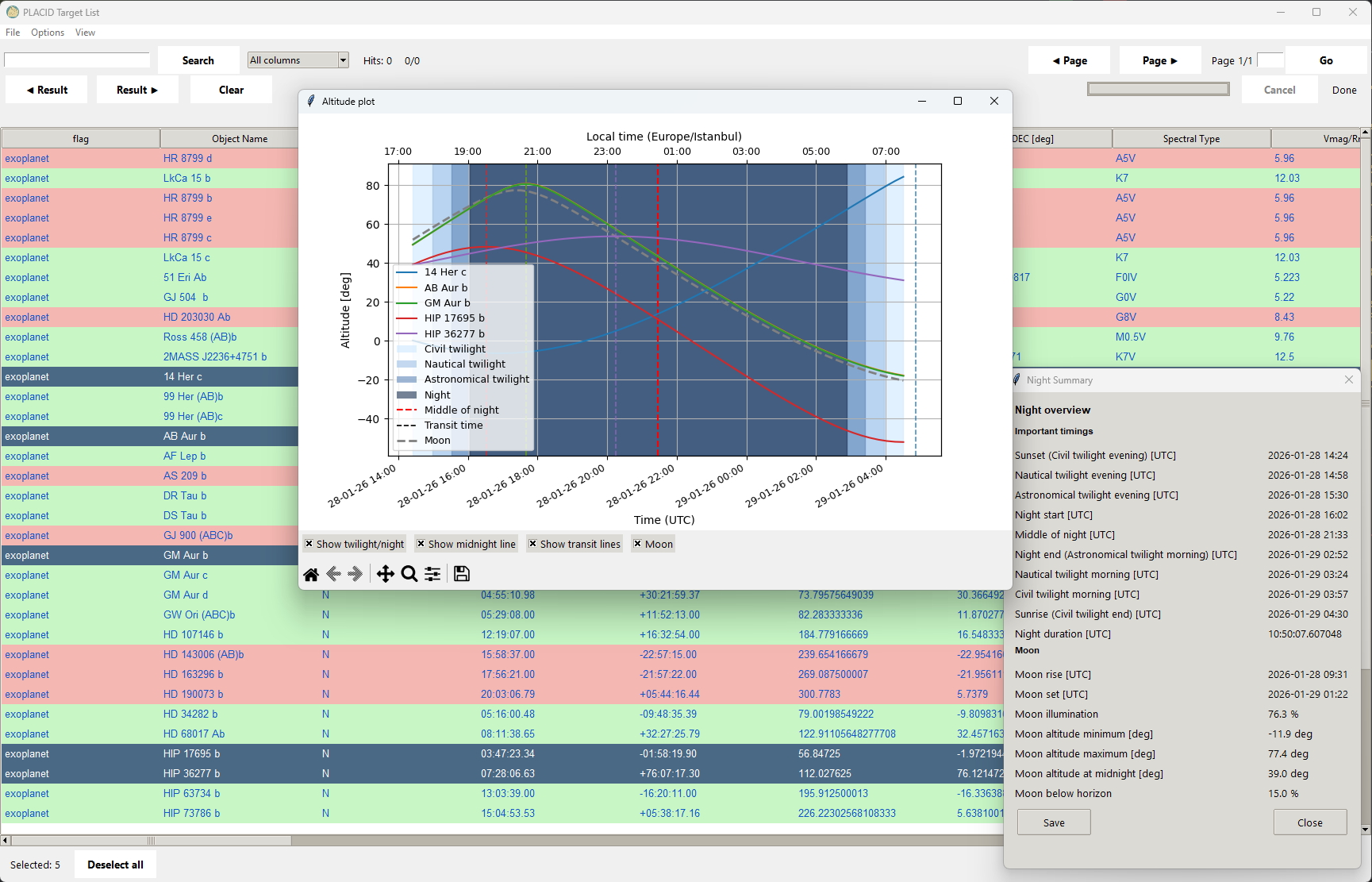}
		\end{tabular}
	\end{center}
   \caption[example] 
   { \label{fig:Target_List_GUI} PLACID Target List tool for preparing observations from the DAG observatory.}
\end{figure}

With the internal alignment of PLACID successfully completed and the normalized intensity levels even outperforming the factory acceptance results, PLACID is more than ready for on-sky commissioning. As of now, while TROIA closed loop tests are ongoing, we are on track to be on-sky in August or September 2026. According to measurements in the last years (from 2023 to 2026), the best month for observations seems to be August with around $\sim$90\% clear night fraction. Also, early fall seems quite useful in terms of clear nights. The median seeing of the DAG site was measured to be 1.16" during the aforementioned period (see Table \ref{tab:DAG}) \cite{Author_InPrep}. \\
Furthermore, the PLACID data reduction software for single stars, the PLACID GUI for single and binary stars, the discovery space and even a target list observation tool are ready (see Figure \ref{fig:Target_List_GUI}). This latter tool facilitates creating own lists and categorizing targets, computing the observability throughout the night for given observation time, computing observability plots, providing rates of field rotation at zenith, information on the selected night (e.g. moon phase, position etc.), and more.

Once on-sky, PLACID will be the first coronagraph of its kind and the only one available for scientific observations, using a programmable focal plane mask. This also makes it the best option to observe binaries with a coronagraph to this date, with many exciting pathways to potential exoplanet, brown dwarf or disk discoveries ahead. Furthermore, PLACID is capable of easily acting as a tool for direct imaging follow-ups for the TESS or PLATO transit candidates or for astrometric targets from GAIA. \\
Additionally, PLACID is on a fast track to implement and test time-domain CDI by changing the phase shift of the focal plane mask locally. In addition, PLACID will be able to compensate for NCPAs using phase-shifting Zernike wavefront sensing with the SLM and the pupil imaging mode of DIRAC, feeding the correction map back towards the TROIA DM, typically as a daytime procedure (or before high-priority science targets).


\newpage

\acknowledgements 
 
The RACE-GO project has received funding from the Swiss State Secretariat for Education, Research and Innovation (SERI), under grant M822.00084 within the framework of the replacement scheme for ERC 2021 Consolidator Grants, following the discontinued participation of Switzerland to Horizon Europe. Part of this work has been carried out within the framework of the National Centre of Competence in Research PlanetS supported by the Swiss National Science Foundation under grants 51NF40 182901 and 51NF40 205606.

\bibliography{PLACID_Comissioning_Status} 

@INPROCEEDINGS{2022SPIE12185E..1WK,
       author = {{Keskin}, Onur and {Jolissaint}, Laurent and {Bouxin}, Audrey and {Yesilyaprak}, Cahit},
        title = "{TROIA adaptive optics system for DAG telescope: assembly and laboratory performance prior to on-sky assessment}",
    booktitle = {Adaptive Optics Systems VIII},
         year = 2022,
       editor = {{Schreiber}, Laura and {Schmidt}, Dirk and {Vernet}, Elise},
       series = {Society of Photo-Optical Instrumentation Engineers (SPIE) Conference Series},
       volume = {12185},
        month = aug,
          eid = {121851W},
        pages = {121851W},
          doi = {10.1117/12.2630893},
       adsurl = {https://ui.adsabs.harvard.edu/abs/2022SPIE12185E..1WK}
}

@INPROCEEDINGS{2018SPIE10706E..2NK,
       author = {{K{\"u}hn}, Jonas and {Patapis}, Polychronis and {Lu}, Xin and {Arikan}, Marcel},
        title = "{SLM-based digital adaptive coronagraphy: current status and capabilities}",
    booktitle = {Advances in Optical and Mechanical Technologies for Telescopes and Instrumentation III},
         year = 2018,
       editor = {{Navarro}, Ram{\'o}n and {Geyl}, Roland},
       series = {Society of Photo-Optical Instrumentation Engineers (SPIE) Conference Series},
       volume = {10706},
        month = jul,
          eid = {107062N},
        pages = {107062N},
          doi = {10.1117/12.2312554},
archivePrefix = {arXiv},
       eprint = {1808.00589},
 primaryClass = {astro-ph.IM},
       adsurl = {https://ui.adsabs.harvard.edu/abs/2018SPIE10706E..2NK}
}

@INPROCEEDINGS{2018SPIE10700E..2JY,
       author = {{Ye{\c{s}}ilyaprak}, Cahit and {Keskin}, Onur},
        title = "{Eastern Anatolia Observatory (DAG): recent developments 2017}",
    booktitle = {Ground-based and Airborne Telescopes VII},
         year = 2018,
       editor = {{Marshall}, Heather K. and {Spyromilio}, Jason},
       series = {Society of Photo-Optical Instrumentation Engineers (SPIE) Conference Series},
       volume = {10700},
        month = jul,
          eid = {107002J},
        pages = {107002J},
          doi = {10.1117/12.2313136},
       adsurl = {https://ui.adsabs.harvard.edu/abs/2018SPIE10700E..2JY}
}

@INPROCEEDINGS{2022SPIE12184E..40Z,
       author = {{Zhelem}, Ross and {Content}, Robert and {Churilov}, Vladimir and {Kripak}, Yevgen and {Waller}, Lew and {Case}, Scott and {Mali}, Slavko and {Muller}, Rolf and {Gonzalez}, Mario and {Adams}, David and {Binos}, Nick and {Chin}, Timothy and {Farrell}, Tony and {Klauser}, Urs and {Kondrat}, Yuriy and {Kunwar}, Nirmala and {Lawrence}, Jon. and {Lorente}, Nuria and {Luo}, Summer and {Mcdonald}, Erica and {McGregor}, Helen and {Nichani}, Vijay and {Pai}, Naveen and {Vuong}, Minh and {Zahoor}, Jahanzeb and {Zheng}, Jessica and {Norris}, Barnaby and {Bryant}, Julia and {Vaccarella}, Annino and {Herrald}, Nick and {Gilbert}, James and {Yesilyaprak}, Cahit and {Gucsav}, Bulent and {Coker}, Deniz and {Keskin}, Onur and {Jolissaint}, Laurent},
        title = "{Design of the near infrared camera DIRAC for East Anatolia Observatory}",
    booktitle = {Ground-based and Airborne Instrumentation for Astronomy IX},
         year = 2022,
       editor = {{Evans}, Christopher J. and {Bryant}, Julia J. and {Motohara}, Kentaro},
       series = {Society of Photo-Optical Instrumentation Engineers (SPIE) Conference Series},
       volume = {12184},
        month = aug,
          eid = {1218440},
        pages = {1218440},
          doi = {10.1117/12.2629716},
       adsurl = {https://ui.adsabs.harvard.edu/abs/2022SPIE12184E..40Z}
}

@inproceedings{10.1117/12.3018836,
author = {Ruben Tandon and Liurong Lin and Axel Potier and Laurent Jolissaint and Audrey Baur and {\"O}zt{\"u}rk {\c{C}}etni, Derya and Jonas G. K{\"u}hn},
title = {{Discovery space and science with the PLACID stellar coronagraph}},
volume = {13096},
booktitle = {Ground-based and Airborne Instrumentation for Astronomy X},
editor = {Julia J. Bryant and Kentaro Motohara and Jo{\"e}l R. D. Vernet},
organization = {International Society for Optics and Photonics},
publisher = {SPIE},
pages = {130963F},
year = {2024},
doi = {10.1117/12.3018836},
URL = {https://doi.org/10.1117/12.3018836}
}

@INPROCEEDINGS{2025SPIE13627E..0ZK,
       author = {{K{\"u}hn}, Jonas G. and {Tandon}, Ruben and {Marquis}, Lucas and {Lin}, Liurong and {{\"O}zt{\"u}rk {\c{C}}etni}, Derya and {Manurung}, Iljadin and {Potier}, Axel and {Jolissaint}, Laurent and {Baur}, Audrey and {Piazza}, Daniele and {Br{\"a}ndli}, Mathias and {Rieder}, Martin},
        title = "{The programmable liquid-crystal active coronagraphic imager for the 4-m DAG telescope (PLACID) instrument: installation and commissioning update}",
    booktitle = {Society of Photo-Optical Instrumentation Engineers (SPIE) Conference Series},
         year = 2025,
       series = {Society of Photo-Optical Instrumentation Engineers (SPIE) Conference Series},
       volume = {13627},
        month = sep,
          eid = {136270Z},
        pages = {136270Z},
          doi = {10.1117/12.3063059},
       adsurl = {https://ui.adsabs.harvard.edu/abs/2025SPIE13627E..0ZK}
}

@unpublished{Author_InPrep,
  author    = {Tu\u{g}rul Tezcan, Cihan},
  title     = {Do\u{g}u Anadolu G\"ozlemevi (DAG) Site Characterization},
  note      = {In preparation (private communication)}
}

@ARTICLE{2001PASP..113..436B,
       author = {{Bourget}, P. and {Veiga}, C.~H. and {Vieira Martins}, R.},
        title = "{A Coronagraph with a Variable-Diameter Occulting Disk}",
      journal = {\pasp},
         year = 2001,
        month = apr,
       volume = {113},
       number = {782},
        pages = {436-438},
          doi = {10.1086/319546},
       adsurl = {https://ui.adsabs.harvard.edu/abs/2001PASP..113..436B}
}

@inproceedings{10.1117/12.2232660,
author = {Jonas K{\"u}hn and Polychronis Patapis},
title = {{Digital adaptive coronagraphy using SLMs: promising prospects of a novel approach, including high-contrast imaging of multiple stars systems}},
volume = {9912},
booktitle = {Advances in Optical and Mechanical Technologies for Telescopes and Instrumentation II},
editor = {Ram{\'o}n Navarro and James H. Burge},
organization = {International Society for Optics and Photonics},
publisher = {SPIE},
pages = {99122M},
year = {2016},
doi = {10.1117/12.2232660},
URL = {https://doi.org/10.1117/12.2232660}
}

@ARTICLE{2014LSA.....3.e213Z,
       author = {{Zhang}, Zichen and {You}, Zheng and {Chu}, Daping},
        title = "{Fundamentals of phase-only liquid crystal on silicon (LCOS) devices}",
      journal = {Light: Science \& Applications},
         year = 2014,
        month = oct,
       volume = {3},
       number = {10},
        pages = {e213-e213},
          doi = {10.1038/lsa.2014.94},
       adsurl = {https://ui.adsabs.harvard.edu/abs/2014LSA.....3.e213Z}
}

@ARTICLE{2022ApOpt..61.9000K,
       author = {{K{\"u}hn}, Jonas G. and {Patapis}, Polychronis},
        title = "{Active focal-plane coronagraphy with liquid-crystal spatial-light modulators: broadband contrast performance in the visible}",
      journal = {\ao},
         year = 2022,
        month = oct,
       volume = {61},
       number = {30},
        pages = {9000},
          doi = {10.1364/AO.467802},
archivePrefix = {arXiv},
       eprint = {2210.14000},
 primaryClass = {astro-ph.IM},
       adsurl = {https://ui.adsabs.harvard.edu/abs/2022ApOpt..61.9000K}
}

@INPROCEEDINGS{2018SPIE10702E..42K,
       author = {{K{\"u}hn}, Jonas and {Daemgen}, Sebastian and {Wang}, Ji and {Morales}, Farisa and {Bottom}, Michael and {Serabyn}, Eugene and {Shelton}, Jean C. and {Delorme}, Jacques-Robert and {Tinyanont}, Samaporn},
        title = "{High-contrast imaging of tight resolved binaries with two vector vortex coronagraphs in cascade with the Palomar SDC instrument}",
    booktitle = {Ground-based and Airborne Instrumentation for Astronomy VII},
         year = 2018,
       editor = {{Evans}, Christopher J. and {Simard}, Luc and {Takami}, Hideki},
       series = {Society of Photo-Optical Instrumentation Engineers (SPIE) Conference Series},
       volume = {10702},
        month = jul,
          eid = {1070242},
        pages = {1070242},
          doi = {10.1117/12.2313448},
archivePrefix = {arXiv},
       eprint = {1808.00585},
 primaryClass = {astro-ph.IM},
       adsurl = {https://ui.adsabs.harvard.edu/abs/2018SPIE10702E..42K}
}

@ARTICLE{2006ApJ...641..556M,
       author = {{Marois}, Christian and {Lafreni{\`e}re}, David and {Doyon}, Ren{\'e} and {Macintosh}, Bruce and {Nadeau}, Daniel},
        title = "{Angular Differential Imaging: A Powerful High-Contrast Imaging Technique}",
      journal = {\apj},
         year = 2006,
        month = apr,
       volume = {641},
       number = {1},
        pages = {556-564},
          doi = {10.1086/500401},
archivePrefix = {arXiv},
       eprint = {astro-ph/0512335},
 primaryClass = {astro-ph},
       adsurl = {https://ui.adsabs.harvard.edu/abs/2006ApJ...641..556M}
}

@ARTICLE{2016A&A...592A..79N,
       author = {{N'Diaye}, M. and {Vigan}, A. and {Dohlen}, K. and {Sauvage}, J.-F. and {Caillat}, A. and {Costille}, A. and {Girard}, J.~H.~V. and {Beuzit}, J.-L. and {Fusco}, T. and {Blanchard}, P. and {Le Merrer}, J. and {Le Mignant}, D. and {Madec}, F. and {Moreaux}, G. and {Mouillet}, D. and {Puget}, P. and {Zins}, G.},
        title = "{Calibration of quasi-static aberrations in exoplanet direct-imaging instruments with a Zernike phase-mask sensor. II. Concept validation with ZELDA on VLT/SPHERE}",
      journal = {\aap},
         year = 2016,
        month = aug,
       volume = {592},
          eid = {A79},
        pages = {A79},
          doi = {10.1051/0004-6361/201628624},
archivePrefix = {arXiv},
       eprint = {1606.01895},
 primaryClass = {astro-ph.EP},
       adsurl = {https://ui.adsabs.harvard.edu/abs/2016A&A...592A..79N}
}

@INPROCEEDINGS{2012SPIE.8447E..2KW,
       author = {{Wallace}, J. Kent and {Crawford}, Sam and {Loya}, Frank and {Moore}, James},
        title = "{A phase-shifting Zernike wavefront sensor for the Palomar P3K adaptive optics system}",
    booktitle = {Adaptive Optics Systems III},
         year = 2012,
       editor = {{Ellerbroek}, Brent L. and {Marchetti}, Enrico and {V{\'e}ran}, Jean-Pierre},
       series = {Society of Photo-Optical Instrumentation Engineers (SPIE) Conference Series},
       volume = {8447},
        month = jul,
          eid = {84472K},
        pages = {84472K},
          doi = {10.1117/12.927041},
       adsurl = {https://ui.adsabs.harvard.edu/abs/2012SPIE.8447E..2KW}
}

@ARTICLE{2012MNRAS.427..948A,
       author = {{Amara}, Adam and {Quanz}, Sascha P.},
        title = "{PYNPOINT: an image processing package for finding exoplanets}",
      journal = {\mnras},
         year = 2012,
        month = dec,
       volume = {427},
       number = {2},
        pages = {948-955},
          doi = {10.1111/j.1365-2966.2012.21918.x},
archivePrefix = {arXiv},
       eprint = {1207.6637},
 primaryClass = {astro-ph.IM},
       adsurl = {https://ui.adsabs.harvard.edu/abs/2012MNRAS.427..948A}
}

@ARTICLE{1997PASP..109..815R,
       author = {{Roddier}, F. and {Roddier}, C.},
        title = "{Stellar Coronograph with Phase Mask}",
      journal = {\pasp},
         year = 1997,
        month = jul,
       volume = {109},
        pages = {815-820},
          doi = {10.1086/133949},
       adsurl = {https://ui.adsabs.harvard.edu/abs/1997PASP..109..815R}
}

@ARTICLE{2000PASP..112.1479R,
       author = {{Rouan}, D. and {Riaud}, P. and {Boccaletti}, A. and {Cl{\'e}net}, Y. and {Labeyrie}, A.},
        title = "{The Four-Quadrant Phase-Mask Coronagraph. I. Principle}",
      journal = {\pasp},
         year = 2000,
        month = nov,
       volume = {112},
       number = {777},
        pages = {1479-1486},
          doi = {10.1086/317707},
       adsurl = {https://ui.adsabs.harvard.edu/abs/2000PASP..112.1479R}
}

@ARTICLE{2005OptL...30.3308F,
       author = {{Foo}, Gregory and {Palacios}, David M. and {Swartzlander}, Jr., Grover A.},
        title = "{Optical vortex coronagraph}",
      journal = {Optics Letters},
         year = 2005,
        month = dec,
       volume = {30},
       number = {24},
        pages = {3308-3310},
          doi = {10.1364/OL.30.003308},
       adsurl = {https://ui.adsabs.harvard.edu/abs/2005OptL...30.3308F}
}

@ARTICLE{2005ApJ...633.1191M,
       author = {{Mawet}, D. and {Riaud}, P. and {Absil}, O. and {Surdej}, J.},
        title = "{Annular Groove Phase Mask Coronagraph}",
      journal = {\apj},
         year = 2005,
        month = nov,
       volume = {633},
       number = {2},
        pages = {1191-1200},
          doi = {10.1086/462409},
       adsurl = {https://ui.adsabs.harvard.edu/abs/2005ApJ...633.1191M}
}

@ARTICLE{2009OExpr..17.1902M,
       author = {{Mawet}, D. and {Serabyn}, E. and {Liewer}, K. and {Hanot}, Ch. and {McEldowney}, S. and {Shemo}, D. and {O'Brien}, N.},
        title = "{Optical Vectorial Vortex Coronagraphs using Liquid Crystal Polymers: theory, manufacturing and laboratory demonstration}",
      journal = {Optics Express},
         year = 2009,
        month = feb,
       volume = {17},
       number = {3},
        pages = {1902-1918},
          doi = {10.1364/OE.17.001902},
archivePrefix = {arXiv},
       eprint = {0912.0311},
 primaryClass = {astro-ph.IM},
       adsurl = {https://ui.adsabs.harvard.edu/abs/2009OExpr..17.1902M}
}

@ARTICLE{2015JOSAA..32.1276H,
       author = {{H{\'e}nault}, Fran{\c{c}}ois},
        title = "{Strehl ratio: a tool for optimizing optical nulls and singularities}",
      journal = {Journal of the Optical Society of America A},
         year = 2015,
        month = jul,
       volume = {32},
       number = {7},
        pages = {1276},
          doi = {10.1364/JOSAA.32.001276},
       adsurl = {https://ui.adsabs.harvard.edu/abs/2015JOSAA..32.1276H}
}

@ARTICLE{2018OptCo.423..186H,
       author = {{H{\'e}nault}, Fran{\c{c}}ois},
        title = "{Analysis of azimuthal phase mask coronagraphs}",
      journal = {Optics Communications},
         year = 2018,
        month = sep,
       volume = {423},
        pages = {186-199},
          doi = {10.1016/j.optcom.2018.04.020},
archivePrefix = {arXiv},
       eprint = {1805.11994},
 primaryClass = {astro-ph.IM},
       adsurl = {https://ui.adsabs.harvard.edu/abs/2018OptCo.423..186H}
}

@ARTICLE{2010ApJS..190....1R,
       author = {{Raghavan}, Deepak and {McAlister}, Harold A. and {Henry}, Todd J. and {Latham}, David W. and {Marcy}, Geoffrey W. and {Mason}, Brian D. and {Gies}, Douglas R. and {White}, Russel J. and {ten Brummelaar}, Theo A.},
        title = "{A Survey of Stellar Families: Multiplicity of Solar-type Stars}",
      journal = {\apjs},
         year = 2010,
        month = sep,
       volume = {190},
       number = {1},
        pages = {1-42},
          doi = {10.1088/0067-0049/190/1/1},
archivePrefix = {arXiv},
       eprint = {1007.0414},
 primaryClass = {astro-ph.SR},
       adsurl = {https://ui.adsabs.harvard.edu/abs/2010ApJS..190....1R}
}

@ARTICLE{2011PASP..123..333C,
       author = {{Cady}, Eric and {McElwain}, Michael and {Kasdin}, N. Jeremy and {Thalmann}, Christian},
        title = "{A Dual-Mask Coronagraph for Observing Faint Companions to Binary Stars}",
      journal = {\pasp},
         year = 2011,
        month = mar,
       volume = {123},
       number = {901},
        pages = {333},
          doi = {10.1086/659038},
archivePrefix = {arXiv},
       eprint = {1103.3275},
 primaryClass = {astro-ph.IM},
       adsurl = {https://ui.adsabs.harvard.edu/abs/2011PASP..123..333C}
}

@ARTICLE{2018AJ....156..291C,
       author = {{Currie}, Thayne and {Brandt}, Timothy D. and {Uyama}, Taichi and {Nielsen}, Eric L. and {Blunt}, Sarah and {Guyon}, Olivier and {Tamura}, Motohide and {Marois}, Christian and {Mede}, Kyle and {Kuzuhara}, Masayuki and {Groff}, Tyler D. and {Jovanovic}, Nemanja and {Kasdin}, N. Jeremy and {Lozi}, Julien and {Hodapp}, Klaus and {Chilcote}, Jeffrey and {Carson}, Joseph and {Martinache}, Frantz and {Goebel}, Sean and {Grady}, Carol and {McElwain}, Michael and {Akiyama}, Eiji and {Asensio-Torres}, Ruben and {Hayashi}, Masa and {Janson}, Markus and {Knapp}, Gillian R. and {Kwon}, Jungmi and {Nishikawa}, Jun and {Oh}, Daehyeon and {Schlieder}, Joshua and {Serabyn}, Eugene and {Sitko}, Michael and {Skaf}, Nour},
        title = "{SCExAO/CHARIS Near-infrared Direct Imaging, Spectroscopy, and Forward-Modeling of {\ensuremath{\kappa}} And b: A Likely Young, Low-gravity Superjovian Companion}",
      journal = {\aj},
         year = 2018,
        month = dec,
       volume = {156},
       number = {6},
          eid = {291},
        pages = {291},
          doi = {10.3847/1538-3881/aae9ea},
archivePrefix = {arXiv},
       eprint = {1810.09457},
 primaryClass = {astro-ph.EP},
       adsurl = {https://ui.adsabs.harvard.edu/abs/2018AJ....156..291C}
}

@ARTICLE{2025A&A...702A...4G,
       author = {{Godoy}, N. and {Choquet}, E. and {Serabyn}, E. and {M{\^a}lin}, M. and {Tremblin}, P. and {Danielski}, C. and {Lagage}, P.~O. and {Boccaletti}, A. and {Charnay}, B. and {Ressler}, M.~E.},
        title = "{A JWST/MIRI view of {\ensuremath{\kappa}} Andromedae b: Refining its mass, age, and physical parameters}",
      journal = {\aap},
         year = 2025,
        month = oct,
       volume = {702},
          eid = {A4},
        pages = {A4},
          doi = {10.1051/0004-6361/202554652},
archivePrefix = {arXiv},
       eprint = {2509.03624},
 primaryClass = {astro-ph.EP},
       adsurl = {https://ui.adsabs.harvard.edu/abs/2025A&A...702A...4G}
}
\bibliographystyle{spiebib} 

\end{document}